\documentclass[twocolumn]{aastex62}

\usepackage{times}
\usepackage{epsfig}
\usepackage{graphicx}
\usepackage{bm}
\usepackage{natbib}
\usepackage{comment}
\usepackage{xcolor}
\usepackage{enumerate}
\usepackage{bigints}
\usepackage{multirow}
\usepackage{booktabs} 
\usepackage{amsmath}
\usepackage{mathtools}
\usepackage{makecell,tabularx}
\usepackage[normalem]{ulem}
\usepackage{float}
\setcellgapes{3pt}
\graphicspath{{./}{figures/}}

\newcommand{\RoNSNS}{\ensuremath{347_{-256}^{+536}\,\mathrm{Gpc^{-3}\,yr^{-1}}}}

\newcommand{\fracjet}{$52\%$}
\newcommand{\fracoffaxis}{$97\%$}
\newcommand{\fracbkt}{$1\%$}
\newcommand{\frackn}{$78\%$}

\newcommand{\searchhlv}{$1.8^{+2.7}_{-1.3}$}
\newcommand{\searchhlvf}{23\%}
\newcommand{\searchhlvk}{$7.7^{+11.9}_{-5.7}$}
\newcommand{\searchhlvkf}{100\%}

\newcommand{\subhlv}{$13^{+20}_{-9.6}$}

\newcommand{\subhlvk}{$54^{+84}_{-40}$}

\newcommand{\searchJ}{$2.4^{+3.6}_{-1.8}$}
\newcommand{\searchz}{$5.1^{+7.8}_{-3.8}$}
\newcommand{\searchg}{$5.7^{+8.7}_{-4.2}$}
\newcommand{\searchJf}{36\%}
\newcommand{\searchzf}{67\%}
\newcommand{\searchgf}{74\%}

\newcommand{\monJ}{$6.0^{+9.2}_{-4.4}$}
\newcommand{\monz}{$6.0^{+9.2}_{-4.4}$}
\newcommand{\mong}{$6.0^{+9.2}_{-4.4}$}
\newcommand{\monJf}{\frackn}
\newcommand{\monzf}{\frackn}
\newcommand{\mongf}{\frackn}

\newcommand{\subJ}{$3.4^{+5.3}_{-2.5}$}
\newcommand{\subz}{$14^{+20}_{-10.4}$}
\newcommand{\subg}{$21^{+34}_{-15}$}

\newcommand{\searchzadio}{$0.29^{+0.44}_{-0.22}$}
\newcommand{\searchoptic}{$0.06^{+0.09}_{-0.04}$}
\newcommand{\searchx}{$0.32^{+0.51}_{-0.23}$}
\newcommand{\searchzadiof}{4\%}
\newcommand{\searchopticf}{0.8\%}
\newcommand{\searchxf}{4\%}

\newcommand{\monzadio}{$0.78^{+1.21}_{-0.58}$}
\newcommand{\monoptic}{$0.47^{+0.74}_{-0.35}$}
\newcommand{\monx}{$0.57^{+0.89}_{-0.42}$}
\newcommand{\monzadiof}{10\%}
\newcommand{\monopticf}{6\%}
\newcommand{\monxf}{7\%}

\newcommand{\subzadio}{$0.95^{+1.45}_{-0.70}$}
\newcommand{\suboptic}{$0.24^{+0.38}_{-0.18}$}
\newcommand{\subx}{$1.23^{+1.89}_{-0.91}$}

\newcommand{\searchBAT}{$0.03^{+0.04}_{-0.02}$}
\newcommand{\searchGBM}{$0.17^{+0.26}_{-0.13}$}
\newcommand{\searchBATf}{0.4\%}
\newcommand{\searchGBMf}{2\%}

\newcommand{\monBAT}{$0.05^{+0.07}_{-0.04}$}
\newcommand{\monGBM}{$0.31^{+0.48}_{-0.23}$}
\newcommand{\monBATf}{0.6\%}
\newcommand{\monGBMf}{4\%}

\newcommand{\subBAT}{$0.12^{+0.19}_{-0.09}$}
\newcommand{\subGBM}{$0.75^{+1.16}_{-0.55}$}

\begin{document}

\title{\textbf{Multi-messenger observations of binary neutron star mergers in the O4 run}}

\author[0000-0002-7439-4773]{Alberto Colombo}
\affiliation{Università degli Studi di Milano-Bicocca, Dipartimento di Fisica "G. Occhialini", Piazza della Scienza 3, I-20126 Milano (MI), Italy}
\affiliation{INFN – Sezione di Milano-Bicocca, Piazza della Scienza 3, I-20126 Milano (MI), Italy}

\author[0000-0003-4924-7322]{{Om Sharan} Salafia}
\affiliation{Università degli Studi di Milano-Bicocca, Dipartimento di Fisica "G. Occhialini", Piazza della Scienza 3, I-20126 Milano (MI), Italy}
\affiliation{INFN – Sezione di Milano-Bicocca, Piazza della Scienza 3, I-20126 Milano (MI), Italy}
\affiliation{INAF – Osservatorio Astronomico di Brera, via E. Bianchi 46, I-23807 Merate (LC), Italy}

\author[0000-0003-3103-9170]{Francesco Gabrielli}
\affiliation{Scuola Internazionale Superiore di Studi Avanzati, via Bonomea 265, I-34136 Trieste (TS), Italy}

\author[0000-0001-5876-9259]{Giancarlo Ghirlanda}
\affiliation{INAF – Osservatorio Astronomico di Brera, via E. Bianchi 46, I-23807 Merate (LC), Italy}
\affiliation{INFN – Sezione di Milano-Bicocca, Piazza della Scienza 3, I-20126 Milano (MI), Italy}

\author[0000-0002-6947-4023]{Bruno Giacomazzo}
\affiliation{Università degli Studi di Milano-Bicocca, Dipartimento di Fisica "G. Occhialini", Piazza della Scienza 3, I-20126 Milano (MI), Italy}
\affiliation{INFN – Sezione di Milano-Bicocca, Piazza della Scienza 3, I-20126 Milano (MI), Italy}
\affiliation{INAF – Osservatorio Astronomico di Brera, via E. Bianchi 46, I-23807 Merate (LC), Italy}

\author[0000-0002-0936-8237]{Albino Perego}
\affiliation{Università di Trento, Dipartimento di Fisica, Via Sommarive 14, 38123 Trento, Italy}
\affiliation{INFN-TIFPA, Trento Institute for Fundamental Physics and Applications, via Sommarive 14, I-38123 Trento, Italy}

\author{Monica Colpi}
\affiliation{Università degli Studi di Milano-Bicocca, Dipartimento di Fisica "G. Occhialini", Piazza della Scienza 3, I-20126 Milano (MI), Italy}
\affiliation{INFN – Sezione di Milano-Bicocca, Piazza della Scienza 3, I-20126 Milano (MI), Italy}
\affiliation{INAF – Osservatorio Astronomico di Brera, via E. Bianchi 46, I-23807 Merate (LC), Italy}

\begin{abstract}
We present realistic expectations for the number and properties of neutron star binary mergers to be detected as multi-messenger sources during the upcoming fourth observing run (O4) of the LIGO-Virgo-KAGRA gravitational wave (GW) detectors, with the aim of providing guidance for the optimization of observing strategies. Our predictions are based on a population synthesis model which includes the GW signal-to-noise ratio, the kilonova (KN) optical and near-infrared light curves, the relativistic jet gamma-ray burst (GRB) prompt emission peak photon flux, and the afterglow light curves in radio, optical and X-rays. Within our assumptions, the rate of GW events to be confidently detected during O4 is \searchhlvk\ yr$^{-1}$ (calendar year), \frackn\ of which will produce a KN, and a lower \fracjet\ will also produce a relativistic jet. The typical depth of current optical electromagnetic search and follow up strategies is still sufficient to detect most of the KN\ae\ in O4, but only for the first night or two. 
The prospects for detecting relativistic jet emission are not promising. While closer events (within $z\lesssim 0.02$) will likely still have a detectable cocoon shock breakout, most events will have their GRB emission (both prompt and afterglow) missed unless seen under a small viewing angle. This reduces the fraction of events with detectable jets to 2\% (prompt emission, serendipitous) and 10\% (afterglow, deep radio monitoring), corresponding to detection rates of \searchGBM\ and \monzadio\ yr$^{-1}$, respectively.
When considering a GW sub-threshold search triggered by a GRB detection, our predicted rate of joint GW+GRB prompt emission detections increases up to a more promising \subGBM\ yr$^{-1}$.
   
\end{abstract}


\section{Introduction} \label{sec:intro}
The second generation of gravitational wave detectors -- now comprising the Advanced Laser Interferometer Gravitational-Wave Observatory (aLIGO, \citealt{AdvancedLIGO2015}), Advanced Virgo (\citealt{AdvancedVirgo2015}) and, starting with the third observing run O3, KAGRA (\citealt{KAGRA2013}) -- led to a revolution in our capability to listen to the Universe, that started with the discovery of GW150914 \citep{GW150914}, the first compact binary coalescence (CBC) detected in gravitational waves (GWs). 
During the first three observing runs (O1, O2 and O3 -- \citealt{GWTC1,LVC2021,GWTC3}), the  network, operated by the LIGO, Virgo and KAGRA (LVK) Collaborations, detected ninety significant (\textit{p}$_\mathrm{astro} > 0.5$) events comprising signals from merging binary black holes (BHBH, the vast majority), binary neutron stars (NSNS, with only two confident identifications) and even black hole-neutron star (BHNS) coalescences  \citep{LVC2021_2,GWTC3}.
The latter detections, performed during the second part of O3, marked the first ever observation of this new type of sources. So far, electromagnetic (EM) emission was observed only in association to the NSNS merger GW170817 \citep{gw170817em}.
Thanks to Advanced Virgo joining the network shortly before, GW170817 was localized in the sky within an area of 28 deg$^2$ (at 90\% credible level, high-latency -- \citealt{Veitch2015}; \citealt{gw170817em}). Remarkably, the localisation was consistent with that of GRB170817A, a short gamma-ray burst (GRB) detected by Fermi and INTEGRAL \citep{Abbott2017_GRB170817A} two seconds after the GW170817 chirp. Telescopes all over the world soon discovered an intrinsically faint, rapidly evolving optical/near-infrared transient in a nearby galaxy within the GW170817 localisation error box (\citealt{Coulter2017}), which was then spectroscopically classified (\citealt{Pian2017}) as a kilonova (KN), that is, quasi-thermal emission from the expanding ejecta produced during and after the merger, powered by the nuclear decay of heavy elements synthesized by rapid neutron capture. In the second week after the merger an additional, broadband (radio to X-rays), non-thermal source was detected at the same position: after a few months, the debate about the nature of the source was settled by very long baseline interferometry observations \citep{Mooley2018,Ghirlanda2019}, which provided conclusive evidence in support of its interpretation as the afterglow of a relativistic jet seen off-axis.

The O3 observing run did not see any new EM counterpart detection (except for a controversial association to the BHBH merger GW190521, see \citealt{Graham2020}), despite the significant increase in sensitivity. The EM follow up campaigns in response to potentially EM-bright O3 events proved generally difficult, in some cases due to the poor sky localisation of the GW signal (e.g.\ in the case of GW190425, \citealt{GW190425}) or to the relatively large distance (e.g.\ GW190814, for which the non-detection of an EM counterpart did not lead to strong constraints on the progenitor -- see for example \citealt{Ackley2020} -- despite the good localisation and the massive observational effort). 

The next, year-long observing run O4 is currently planned\footnote{\url{https://www.ligo.caltech.edu/news/ligo20211115}} to start in December 2022. The improvements in the sensitivity of the LIGO Hanford and Livingston, Virgo and KAGRA (HLVK) interferometers will let us explore a wider volume of the Universe, with a large predicted increase in the detection rate with respect to O3 \citep{Abbott-rates2020,Petrov2021}. The optimisation of EM follow-up strategies will be fundamental in order to enhance the probability of discovering rapidly fading transients in association to these detections. Indications about the predicted GW and EM properties of the population accessible during O4 would be extremely valuable to this task (see \citealt{Barbieri2020} for an application using the expected kilonova light curve range). 

In this article we present our predictions\footnote{When needed, in this work we assume a $\Lambda$CDM cosmology, with \citet{Planck} parameter, namely $\Omega_\mathrm{M} = 0.3065$, $\Omega_\lambda = 0.6935$, $\Omega_\mathrm{k} = 0.005$, $h = 0.679$. 
Errors due to the uncertainty in cosmological parameters are negligible in comparison to intrinsic rate density uncertainties and modelling systematics.} for the observational appearance of the EM emission associated to NSNS mergers that will be detected during O4, focusing on KN and jet-related emission. 
To this purpose, we built a synthetic population of merging NSNS binaries, with a mass distribution informed by both GW and Galactic NSNS binaries (see Appendix~\ref{apx:mass}), and computed the expected properties of their ejecta and accretion disks through numerical-relativity-informed fitting formulae. Using these properties as inputs, we then computed the observable properties of their associated KN, GRB prompt and GRB afterglow emission through a suite of semi-analytical models, updating the methodology described in \citealt{Barbieri2019}. This allowed us to construct the distributions of the EM observables for O4 GW-detectable events, and to address a number of fundamental questions, such as:
\textit{what is the fraction of NSNS mergers that produce an EM counterpart? Which counterpart is best detected in wide-area surveys or in targeted observations? How diverse is the kilonova emission in terms of brightness and other properties? How long after the merger do we expect the detection of most of the GRB afterglows in the radio, optical and X-ray bands?}

\section{Prospects for EM counterpart search and monitoring in O4} \label{sec:EMO4}

\subsection{Multi-messenger observing scenarios and detection limits}

We consider two representative sets of detection limits (see Table~\ref{tab:det_rates}) based on the typical depth that can be reached during an EM follow up in response to a GW alert. In particular, the `counterpart search' set  is representative of the search for an EM counterpart over the GW localization volume (or of online triggering algorithms in the case of space-based gamma-ray detectors), while the `candidate monitoring' set consists of deeper limits typical of the monitoring of a candidate counterpart with arc-second localisation (or of off-line sub-threshold searches in gamma-ray detector data).
In addition to discussing the expected rates of GW+EM events that exceed (some combinations of) these limits, we also briefly discuss the prospects for joint GW+EM detections in off-line sub-threshold searches in GW data triggered, for example, by a GRB detection by an EM facility \citep[we call this a `sub-threshold GW search', see e.g.][]{Abbott2017_grbsearch}.

\citealt{Abbott-rates2020} predicted an optimistic 90\% credible O4 GW localization area of $33^{+5}_{-5}$ deg$^2$ assuming a HLVK network configuration, while \citealt{Petrov2021} proposed a higher and more realistic value of $1820^{+190}_{-170}$ deg$^2$, considering a different minimum number of detectors and a different minimum signal-to-noise ratio (SNR) threshold (based on O3 public alerts).
Given these expected GW localization areas in O4, optical/infrared counterpart searches covering a significant fraction of the localization probability will be only feasible with large, wide-field telescopes or in a galaxy-targeted approach with smaller facilities. In both cases, the typical realistic depth of EM counterpart search observations is down to 21 -- 22 AB magnitudes in the $J, z, g$ bands \citep[e.g.][]{Coughlin2019,Ackley2020}, in part limited by the availability of deep templates. Radio telescopes with a sufficiently fast survey speed can conduct searches for an EM counterpart over a significant fraction of the GW error box \citep{Dobie2022}, either by means of an unbiased survey of the area or by preferentially targeting galaxies, realistically reaching detection limits of $0.1$ mJy at a representative frequency of 1.4 GHz \citep[e.g.][]{Dobie2021,Alexander2021}; X-ray searches have been attempted with the Neil Gehrlels \textit{Swift} Observatory \citep[][]{Page2020} and typically reached a $10^{-13}\,\mathrm{erg\,cm^{-2}}\,\mathrm{\,s^{-1}\,keV^{-1}}$ limiting flux at 1 keV. Despite not representing technically a search, we include in this category the gamma-ray sky monitoring by \textit{Fermi}/GBM and \textit{Swift}/BAT, with representative 64-ms peak photon flux limits of $4$ and $3.5$ ph cm$^{-2}$ s$^{-1}$ (these limits are based on a visual comparison of the flux distribution predicted by our model with those observed by these instruments, see Figure \ref{fig:lognlogs} in Appendix \ref{apx:prompt}).

Once a promising candidate is localized with arc-second accuracy, longer exposures become feasible, and deeper limits can be reached: our deeper `candidate monitoring' detection limit set assumes a detection to be possible down to 28 AB magnitudes in the $J, z, g$ bands, representative of deep space-based observations or of ground-based ones with large adaptive-optics-equipped facilities \citep[e.g.][]{Lyman2018}; in X-rays down to  $10^{-15}\,\mathrm{erg\,cm^{-2}}\,\mathrm{\,s^{-1}\,keV^{-1}}$ at 1 keV, representative of the limits that can be reached by \textit{Chandra} or XMM-Newton with long ($\gtrsim 10^4$ s) exposures \citep[e.g.][]{Margutti2017,Davanzo2018}; in radio down to $10$ $\mu$Jy, representative of limits that can be reached after hour-long exposures with a large facility such as the Karl Jansky Very Large Array \citep[e.g.][]{Hallinan2017}. We also include in this category the off-line, sub-threshold detection of gamma-ray emission by \textit{Fermi}/GBM and \textit{Swift}/BAT, with a representative flux limit of $1$ ph cm$^{-2}$ s$^{-1}$, for both. 

Based on the above considerations, we defined the set of representative detection limits given in Table~\ref{tab:det_rates}.
For the GW detection, we assumed a network SNR threshold $\mathrm{SNR}_\mathrm{net}\geq 12$ (see next sub-section for the definition) for a confident detection, following \citealt{Abbott2020}. They also assume an SNR threshold  larger than 4 in at least two detectors, but this condition would decrease our rate of about 0.1\%. We also tested the condition that the SNR of each single detector should be at least larger than 5, resulting in a decreasing of the rate lower than 5\%. For the sub-threshold GW search we assume a less stringent $\mathrm{SNR}_\mathrm{net}\geq 6$.

\subsection{GW-EM population model}\label{sec:GWEM model}
Our synthetic cosmological population of NSNS mergers is characterized by power-law chirp mass and mass ratio probability distributions, assumed independent and fitted to currently available observational constraints from both GW-detected and Galactic NSNS binaries (see Appendix \ref{apx:mass}). We assumed a cosmic merger rate density (Appendix \ref{apx:redshift}) obtained by convolving a simple $P(t_\mathrm{d})\propto t_\mathrm{d}^{-1}$ delay time distribution (here $t_\mathrm{d}$ represents the delay between the formation of the binary and its GW-driven merger), with a minimum delay $t_\mathrm{d,min}=50\,\mathrm{Myr}$, with the cosmic star formation rate from \citealt{Madau2014}, and normalized (see Appendix \ref{apx:R0}) to a local rate density $R_0=\RoNSNS$ to self-consistently reproduce the actual number of significant NSNS mergers observed so far \citep{GWTC3} . For each event we computed the expected SNR in the LIGO\footnote{\url{https://www.ligo.caltech.edu/}},  Virgo\footnote{\url{https://www.virgo-gw.eu/}} and KAGRA\footnote{\url{https://gwcenter.icrr.u-tokyo.ac.jp/en/}} detectors with the projected O4 sensitivities\footnote{For LIGO, Virgo and KAGRA we considered, respectively, a target sensitivity of 190 Mpc, 120 Mpc and 25 Mpc (\url{https://dcc-lho.ligo.org/LIGO-T2000012/public}). KAGRA will start with 1 Mpc (\url{https://www.ligo.org/scientists/GWEMalerts.php}), but eliminating this detector from the network would result in a decrease of the rates lower than $0.6\%$, making our assumption negligible.}, adopting the \textit{TaylorF2} approximant from LALSimulation through the software package \texttt{PyCBC} to model the GW signal \citep{Abbott2020}, and computed the network SNR as $\mathrm{SNR_{net}} = \sqrt{\sum\mathrm{SNR}_i^2}$ (where $i$ runs over the detectors in the network and we assumed 80\% duty cycle for each detector, in practice setting each single-detector SNR$_i$ to zero randomly with 20\% probability). For all events in the population we then computed the expected ejecta mass, ejecta average velocity and accretion disk mass using numerical-relativity-informed fitting formulae (\citealt{Barbieri2020,krouger2020,Radice2018_2}) and assuming the SFHo equation of state \citep{Steiner_2013}, which satisfies the current astrophysical constraints \citep[e.g.][]{Miller2019}. This equation of state predicts a maximum non-rotating NS mass of $M_\mathrm{TOV}=2.06\:M_{\odot}$\footnote{This implies that the secondary component of GW190814 \citep{Abbott2020_GW190814} is most likely a black hole.}. We used the results as inputs to compute KN light curves from $0.1$ to $50$ days in the $g$ ($484$ nm central wavelength),
$z$ ($900$ nm) 
and $J$ ($1250$ nm) 
bands, using the multi-component model of \citealt{Perego2017} with updates based on \citealt[][see Appendix \ref{apx:kn} for more details]{Breschi2021}. In cases of mergers with final mass $M_\mathrm{rem}\geq 1.2 M_\mathrm{TOV}$, corresponding to remnants that collapse promptly or after a short-lived hyper-massive neutron star phase to a black hole, we assumed the system to launch a relativistic jet, with an energy set by the mass of the accretion disk and the spin of the remnant (see Appendix \ref{apx:rel_jet}). In cases in which the jet energy exceeded a threshold defined following \citealt{Duffell2018}, we assumed the relativistic jet to be able to break out of the ejecta cloud and produce GRB prompt and afterglow emission (a `successful jet'). In our population, \fracjet\ of the events launch a successful jet, satisfying the current observational constraints on the incidence of jets in NSNS mergers \citep{Salafia2022}. For these cases, we assumed a jet angular structure\footnote{Angular dependence of the jet energy density and bulk Lorentz factor.} inspired by GRB170817A \citep[][see Appendix \ref{apx:rel_jet} for more details]{Ghirlanda2019} and computed afterglow light curves from $0.1$ to $1000$ days in the radio ($1.4$ GHz), optical ($g$ band)\footnote{We do not consider the dust extinction in computing the optical KN and GRB afterglow emission.} and X-rays (1 keV), fixing the interstellar medium density at $n=5\times10^{-3}\,\mathrm{cm^{-3}}$ (the median density in the \citealt{Fong2015} sample) and the afterglow microphysical parameters at $\epsilon_\mathrm{e}=0.1$, $\epsilon_\mathrm{B}=10^{-3.9}$ and $p=2.15$ (representative of GW170817, \citealt{Ghirlanda2019}). 
Given the uncertainty on the detailed physical processes involved in the GRB prompt emission, to compute its properties we adopted a semi-phenomenological model similar to that used in \citealt{Barbieri2019} and \citealt{Salafia2019}, where a constant fraction $\eta_\gamma=0.15$ \citep{Beniamini2016} of the jet energy density at each angle (restricting to regions with a bulk Lorentz factor $\Gamma\geq 10$) is assumed to be radiated in the form of photons with a fixed spectrum in the comoving frame. The observed spectrum was then obtained by integrating the resulting radiation over the jet solid angle, accounting for relativistic beaming. To account for a putative wider-angle cocoon shock breakout component \citep{Gottlieb2018}, for systems observed within a viewing angle $\theta_\mathrm{v}\leq 60^\circ$ we also included an additional emission component whose properties reproduce those observed in GRB170817A \citep{Abbott2017_GRB170817A}, namely a luminosity $L_\mathrm{SB}=10^{47}\,\mathrm{erg/s}$ and a cut-off power-law spectrum with $\nu F_\nu$ peak photon energy $E_\mathrm{p,SB}=185\,\mathrm{keV}$ and low-energy photon index $\alpha=-0.62$. The photon fluxes in the 10-1000 keV (\textit{Fermi}/GBM) and 15-150 keV (\textit{Swift}/BAT) energy bands were then computed assuming a fixed rest-frame duration $T=2\,\mathrm{s}$ for all bursts.  We provide more details on the model in Appendix \ref{apx:prompt}. To compute the final GRB prompt emission detection rates we took into account the limited field of view and duty cycle of {\it Fermi}/GBM and {\it Swift}/BAT by multiplying the resulting rates by 0.60 and 0.11 respectively \citep{Burns2016}.

\begin{table*}[htbp]
\caption{Assumed detection limits and predicted detection rates in our observing scenarios. Below each rate we also report in parentheses the fraction over the total O4 NSNS GW rate (“HLVK O4”). The GW detection limits refer to the $\mathrm{SNR_{net}}$ threshold. Near infrared and optical limiting magnitudes are in the AB system; radio limiting flux densities are in mJy @ 1.4 GHz; X-ray limiting flux densities are in erg cm$^{-2}$ s$^{-1}$ keV$^{-1}$ @ 1 keV; gamma-ray limiting photon fluxes are in photons cm$^{-2}$ s$^{-1}$ in the 15--150 keV (\textit{Swift}/BAT) or 10--1000 keV (\textit{Fermi}/GBM) band. Detection rates are in $\mathrm{yr}^{-1}$. The reported errors, given at the 90\% credible level, stem from the uncertainty on the overall merger rate (hence they cancel out in the fractions), while systematic errors are not included.}
\centering
\begin{tabular}{c|cc|ccc|ccc|cc}

\multicolumn{1}{c|}{\multirow{2}{*}{\textbf{}}} & \multicolumn{2}{c|}{GW} & \multicolumn{3}{c|}{Kilonova + GW O4} & \multicolumn{3}{c|}{GRB Afterglow + GW O4} & \multicolumn{2}{c}{GRB Prompt + GW O4} \\
& HLV O3 & HLVK O4 & \textit{J} & \textit{z} & \textit{g} & Radio & Optical & X-rays & \textit{Swift}/BAT & \textit{Fermi}/GBM \\ \hline
\textbf{Count. Search} & ~ & ~ & ~ & ~ & ~ & ~ &  &  &  & \\
Limit & 12 & 12 & 21 & 22 & 22 & 0.1 & 22 & $10^{-13}$ & $3.5$ & $4$  \\ 

Rate & \searchhlv & \searchhlvk & \searchJ & \searchz & \searchg & \searchzadio & \searchoptic & \searchx & \searchBAT & \searchGBM \\ 
(\% of O4 GW) & (\searchhlvf) & (\searchhlvkf) & (\searchJf) & (\searchzf) & (\searchgf) & (\searchzadiof) & (\searchopticf) & (\searchxf) & (\searchBATf) & (\searchGBMf) \\
\hline
\textbf{Cand. Monitoring} & ~ & ~ & ~ & ~ & ~ & ~ &  &  &  & \\
Limit & / & / & 28 & 28 & 28 & 0.01 & 28 & $10^{-15}$ & $1$  & $1$  \\ 
Rate & / & / & \monJ & \monz & \mong & \monzadio & \monoptic & \monx & \monBAT & \monGBM \\   
(\% of O4 GW) & / & / & (\monJf) & (\monzf) & (\mongf) & (\monzadiof) & (\monopticf) & (\monxf) & (\monBATf) & (\monGBMf) \\
\hline
\textbf{GW subthreshold} & ~ & ~ & ~ & ~ & ~ & ~ &  &  &  & \\
Limit & 6 & 6 & 21 & 22 & 22 & 0.1 & 22 & $10^{-13}$ & $3.5$  & $4$  \\ 
Rate & \subhlv & \subhlvk & \subJ & \subz & \subg & \subzadio & \suboptic & \subx & \subBAT & \subGBM \\   

\end{tabular}
\label{tab:det_rates}
\end{table*}

\begin{figure*}
    \centering
    \includegraphics[width=0.99\textwidth]{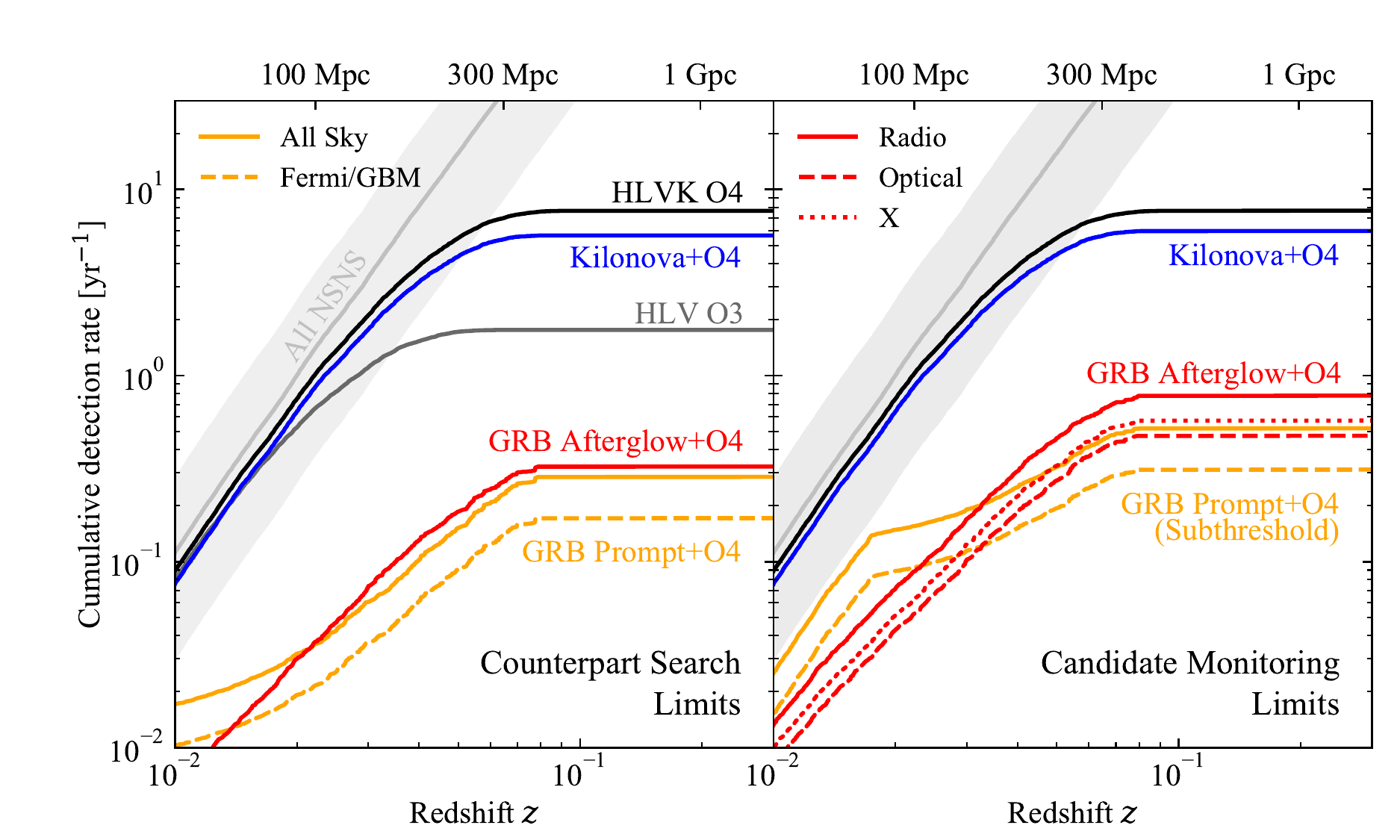}
    \caption{Cumulative multi-messenger detection rates as a function of redshift (luminosity distance) for our NSNS population. The left-hand panel assumes the `counterpart search' detection limits, representative of a search for an EM counterpart over the GW localization volume (see Tab.~\ref{tab:det_rates}). The light grey line (“All NSNS”) represents the intrinsic merger rate in a cumulative form, with the grey band showing its assumed uncertainty \citep{GWTC2_1}, which propagates as a constant relative error contribution to all the other rates shown in the figure. The black (“HLVK O4”) and grey (“HLV O3”) lines are the cumulative GW detection rates (events per year with network SNR $\geq 12$, accounting for the single detector duty cycles) in O4 and O3. The blue (“Kilonova+O4”), red (“GRB Afterglow+O4”) and orange (“GRB Prompt+O4”) lines are the cumulative detection rates for the joint detection of GW and a KN, GRB afterglow or GRB prompt in O4 (in at least one of the considered bands, all-sky except for the dashed line, which accounts for the \textit{Fermi}/GBM duty cycle and field of view). The right-hand panel assumes deeper detection limits (see Tab.~\ref{tab:det_rates}) representative of the monitoring of a well-localized candidate (and a sub-threshold search for the GRB prompt). For the GRB afterglow we show separately the radio, optical and X-ray band detection rates (solid, dashed and dotted, respectively). }
    \label{fig:detection}
\end{figure*}

\subsection{Detection rates with the `counterpart search' limit set}
In the left-hand panel of Figure \ref{fig:detection} we show our predictions for the EM counterpart search scenario in O4, assuming the `counterpart search'  limits set. The light grey line (“All NSNS”) represents the intrinsic cumulative merger rate, with the underlying light grey band showing its uncertainty (Poissonian uncertainty on the rate density normalization assuming our mass distribution, see Appendix~\ref{apx:R0}), which propagates as a constant relative error contribution to all the other rates shown in the figure. The black line (“HLVK O4”) is our prediction for the cumulative detection rate of NSNS mergers by the GW detector network in O4. For comparison we also show, with a dark grey line, the rate assuming the HLV O3 configuration network and duty cycle \citep{LVC2021}. The blue, red and orange lines are the all-sky cumulative rates for the joint detection of GW and KN\ae\  (“Kilonova+O4”), GW and GRB afterglows (“GRB Afterglows+O4”), GW and GRB prompt emission (“GRB Prompt+O4”), respectively. For the latter we show the rate for a GRB detection by \textit{Fermi}/GBM (dashed line) and, for comparison, the rate of a putative detector with the same sensitivity, but with an all-sky field of view and a 100\% duty cycle (solid line). The result for \textit{Swift}/BAT is reported in Table \ref{tab:det_rates}. The redshift (or luminosity distance) values at which the curves saturate clearly show that the horizons are currently set by the GW detection.

We find that, in the O4 run, NSNS merger GW signals will be detectable out to $\sim 300$ Mpc ($z \sim 0.07$), with a detection rate of \searchhlvk events per calendar year. Joint GW+EM detection rates for the various counterparts considered are reported in Table~\ref{tab:det_rates}. These rates show that the vast majority of KN\ae\  associated to O4 events will be brighter than the assumed limits at peak, and therefore in principle within the reach of current EM counterpart search facilities and strategies. As shown in Figure~\ref{fig:kn} and detailed in sec.~\ref{sec:kn}, though, the extremely fast evolution of these sources will make their actual identification very challenging, and will require a coordinated global effort and the use of large facilities. Our predicted joint GW and GRB rates for EM searches are instead much lower (\searchx  $\;\mathrm{yr}^{-1}$ for the GRB afterglow and \searchGBM  $\;\mathrm{yr}^{-1}$ for the GRB prompt\footnote{These values can be scaled for different detection limits using Figure \ref{fig:det_lim} in Appendix \ref{apx:det_lim}}), and they reflect the faintness of these components for the considered flux thresholds, mainly because of the large abundance (\fracoffaxis) of off-axis jets (i.e.\ with $\theta_\mathrm{v}\geq 2\theta_\mathrm{c}$, where $\theta_\mathrm{c}$ is the core angle as defined in Appendix \ref{apx:rel_jet}).

\subsection{Detection rates with the `candidate monitoring' limit set}
In the right-hand panel of Figure \ref{fig:detection} we show the results for the scenario simulating the monitoring of a well-localized candidate and the GRB prompt sub-threshold detection, assuming the `candidate monitoring' limits set. These rates represent the hypothetical maximum dection rates that can be achieved in the limiting situation in which all events are localised to arc-second accuracy, allowing for observations as deep as the assumed limits. The KN rate in this panel is therefore shown mostly for reference, as the most likely scenario is one in which the arc-second localisation is obtained through the identification of the KN in a shallower wide-area search. Still, given that all jet-producing events in our population also produce a KN, and given that almost all our KN\ae\ exceed the `couterpart search' limit set, the rates reported for the afterglow in this panel do represent actual achievable rates. 

The light grey and black lines in the panel are the same as the left panel. The blue and red lines are the all-sky cumulative detection rates for the Kilonova+O4 and GRB Afterglow+O4 detectable sources with this limit set. For the latter emission we report individually the rates of events exceeding the radio, optical and X-ray detection thresholds (solid, dashed and dotted line, respectively), showing radio to be the most promising band for the detection of a faint GRB afterglow counterpart. 

In this panel we also show with orange lines the rates of joint GW+GRB detections assuming a detection threshold (see Table~\ref{tab:det_rates}) representative of an off-line sub-threshold search in the gamma-ray detector data. 

The fact that the deeper optical and infrared limits do not increase significantly the KN detection rate reflects the fact that the majority of KN\ae\ associated to O4 events in our population are already brighter (at peak) than the limits adopted in the search scenario. As far as the GRB afterglow is concerned, we find that the deeper limits allow to increase the detection rate in the radio, optical and X-ray bands by factors of $\sim 3$, 8.5 and 2, respectively, with the highest detection rate in the radio, reaching \monzadio $\;\mathrm{yr}^{-1}$. Also for the GRB prompt sub-threshold detection we find a small increase in the rates up to \monGBM $\;\mathrm{yr}^{-1}$.
All detection rates are reported in Table \ref{tab:det_rates}.

\subsection{Sub-threshold GW search in response to an external EM trigger}

The bottom group of rows in Table \ref{tab:det_rates} report the detection rates predicted by our model adopting a lower GW detection threshold $\mathrm{SNR_\mathrm{net}}\geq 6$, which we take as representative of a sub-threshold GW search for events coincident with an external EM trigger. The most relevant external trigger, in our context, is a GRB, as it allows for the search to focus on a short time interval and on a relatively small sky area, therefore increasing significantly the sensitivity with respect to an all-sky, all-time search \citep{Abbott2017_grbsearch}. Thanks to the expanded GW horizon in the sub-threshold search, the rate of joint GRB+GW detections increases to a more promising \subGBM\ $\mathrm{yr}^{-1}$ (for \textit{Fermi/GBM}), which is in good agreement with the rate predicted by the LVK Collaboration for the same kind of search \citep{Abbott2021_GRB} and would mean a relatively high chance of a new GRB-NSNS association. Sub-threshold searches may in principle be conducted also in response to the EM detection of a KN or GRB orphan afterglow candidate: for that reason, we also report the joint GW+KN and GW+afterglow rates in the table, but we caution that these are not representative of a real expected rate, as the serendipitous discovery of KN\ae\  and orphan GRB afterglows in current all-sky surveys is hampered by limited cadence, depth, and availability of time at large facilities for spectroscopic classification of candidates.

\section{EM Properties}

In the following section, we characterize the EM properties of the GW-detectable (with $\mathrm{SNR_\mathrm{net}}\geq 12$) NSNS mergers in our population. Our purpose is mainly that of informing EM follow up strategies, by constructing expected distributions of source brightness at various times and frequencies and for different EM counterparts.

\subsection{Kilonova} \label{sec:kn}
\begin{figure*}
    \centering
    \includegraphics[width=\textwidth]{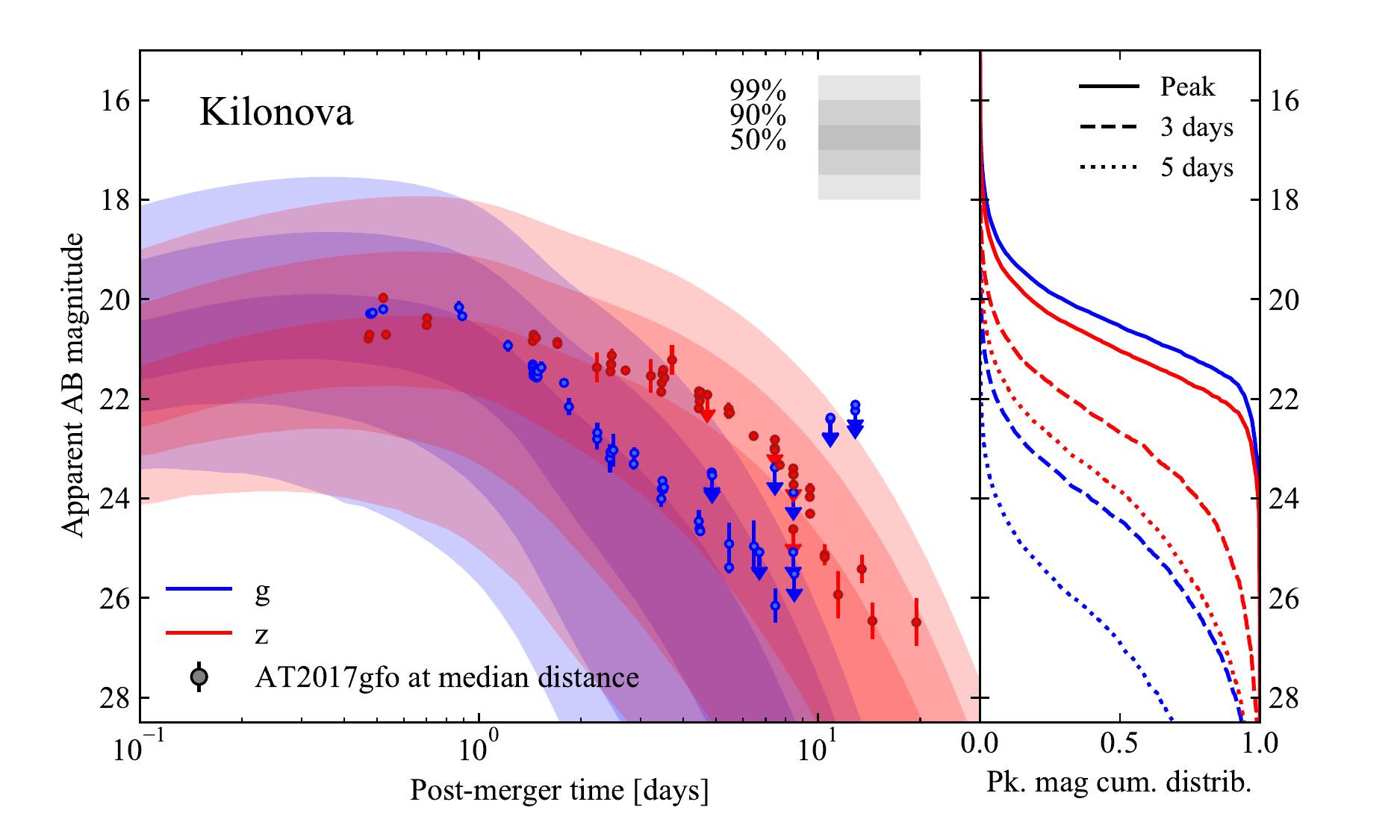}
    \caption{Distribution of O4 KN optical and near-infrared magnitude as a function of time. The left-hand panel shows the apparent AB magnitude versus post-merger time for our simulated KN light curves, restricting to O4 GW-detectable sources. The shaded regions contain $50\%$, $90\%$ and $99\%$ of the KN light curves. Blue and red colors refer respectively to the g (484 nm) and z (900 nm) band. Colored circles show extinction-corrected AT2017gfo data rescaled to the median distance of our population ($\sim181$ Mpc)}.
    The right-hand panel shows the cumulative distributions of apparent magnitude at peak, at 3 days and at 5 days after the merger (solid, dashed and dotted lines, respectively).
    \label{fig:kn}
\end{figure*}

In Figure \ref{fig:kn} we show the time evolution of the distribution of KN brightness for binaries in our population that are GW-detectable in O4. In particular, in the left-hand panel we show the bands that contain $50\%$, $90\%$ and $99\%$ of the light curves at each time. Blue and red colors refer to the $g$ and $z$ band, respectively (we show the corresponding result for the $J$ band in Figure \ref{fig:kn_gJ} in Appendix \ref{apx:kn}).  When scaled to the median distance ($\sim181$ Mpc) of these events, AT2017gfo (colored circles) lies at the top of the $50\%$ band, showing that our assumptions are conservative in that they predict KN\ae\  that are on average slightly dimmer than AT2017gfo, but with a similar temporal evolution. While the peaks of these KN\ae\  span a relatively wide apparent magnitude range $17-24$, $50\%$ are concentrated in the relatively narrow interval $20-22$. In the right-hand panel we show the cumulative apparent magnitude distributions at peak (solid line) and also 3 and 5 days after the merger (dashed and dotted lines), which clearly display the very rapid evolution, especially in the $g$ band. 

The detection of the KN in the \textit{g} and \textit{z} band seems particularly probable (74\% and 63\% of GW events, respectively) for current all-sky EM campaign. However, the rapid evolution, underlined by Figure \ref{fig:kn}, suggests that the observation should take place within the first night for the g band and within about two nights for the \textit{z} band. While the J band, even if it evolves more slowly, is too faint to be detected with current all-sky facilities.

\subsection{GRB Afterglow} \label{sec:after}

\begin{figure}[t]
    \centering
    \includegraphics[width=\columnwidth]{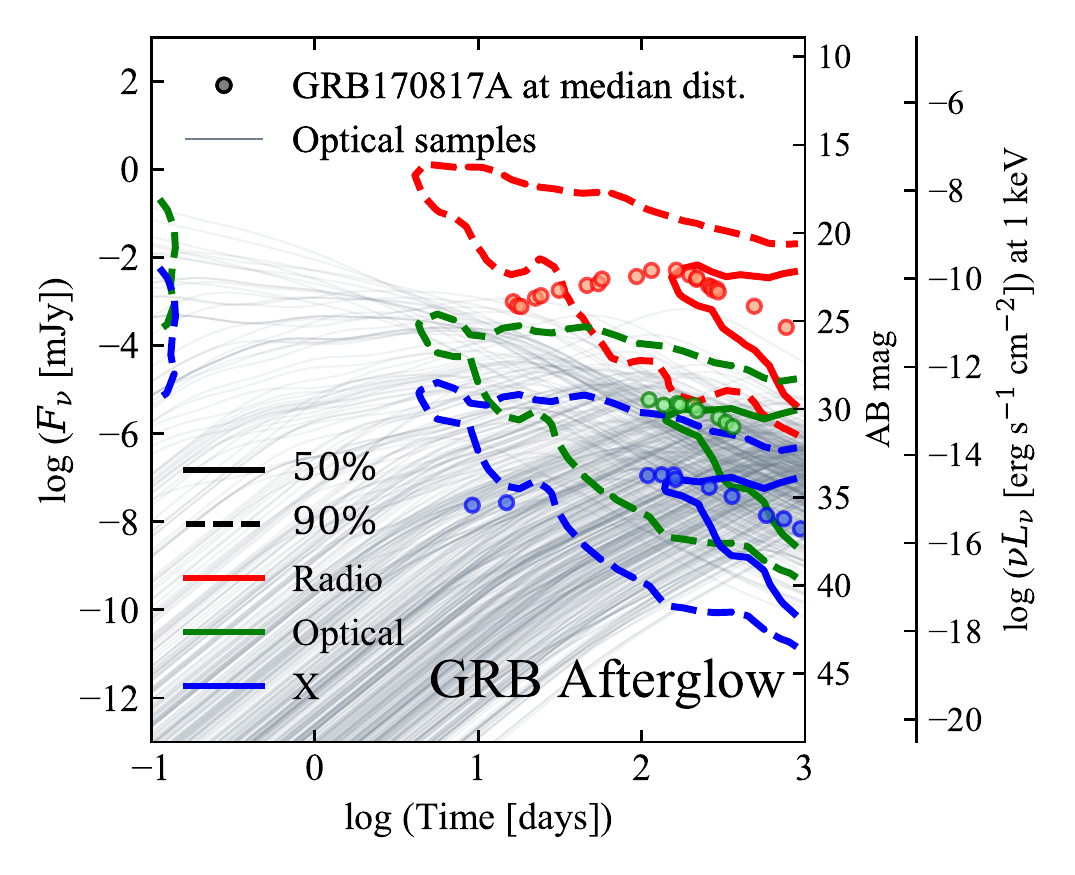}
    \caption{$F_\nu$, AB magnitude and $\nu L_\nu$ versus time for the GRB afterglow light curves associated to O4-detectable sources in our population. Solid and dashed contours contain $50\%$ and $90\%$ of the peaks, respectively. Red, green and blue colors indicate the radio ($1.4 \times 10^9$Hz), optical ($4.8 \times 10^{14}$Hz), X-ray ($2.4 \times 10^{17}$Hz) bands, respectively. The colored circles are the observed data of GRB170817A \citep{Makhathini2021} at the median distance of our population. The grey lines in the background are 500 randomly sampled optical light curves.}
    \label{fig:after}
\end{figure}

In Figure \ref{fig:after} we show the properties of GRB afterglows associated to GW-detectable binaries in our population by showing the contours containing $50\%$ (solid lines) and $90\%$ (dashed lines) of GRB afterglow peaks on the $(F_\nu, t)$ plane. We also report on the right the apparent AB magnitude and $\nu L_\nu$ at 1 keV, where $L_\nu = 4\pi{d_L}^2F_\nu/(1+z)$ is the specific luminosity, $\nu$ is the observer frequency and $t$ is the observer time. The red, green and blue colors refer to our radio, optical and X-ray bands, respectively. Most peak times are at  $\gtrsim 10^2$ days (we note that we restricted the light curve computation between $10^{-1}$ and $10^3$ rest-frame days), with a tail at shorter peak times. We also show 500 randomly sampled optical light curves (thin grey lines) in the background, to help visualizing the underlying light curve behavior. For comparison, we also show GRB170817A data \citep[][small circles]{Makhathini2021} at the median distance of our population ($\sim181$ Mpc), whose peak lies within the 50\% contours in all three bands.

These results stem from the strong dependence of the GRB afterglow light curve on the viewing angle, combined with the GW-detection-induced bias on the viewing angle distribution (which skews the distribution towards smaller viewing angles with respect to the isotropic case, with a peak at $\sim 30^\circ$ -- \citealt{Schutz2011}). This places the majority of the peaks months to years after the GW event, with a small sub-sample peaking at early times ($\sim$hours) in the optical and X-rays, producing very bright emission, thanks to a smaller viewing angle. 

The GRB afterglow properties highlighted in Figure \ref{fig:after} and the low rates shown in Table \ref{tab:det_rates}, suggest that the preferred candidate for an all-sky observing campaign is the KN, at least in the first days after the GW event. Once the KN is detected, it seems convenient to wait weeks or months  after the event, to proceed in search of a GRB afterglow with deeper detection limits.

\subsection{GRB Prompt} \label{sec:grbprompt}

\begin{figure}[t]
    \centering
    \includegraphics[width=\columnwidth]{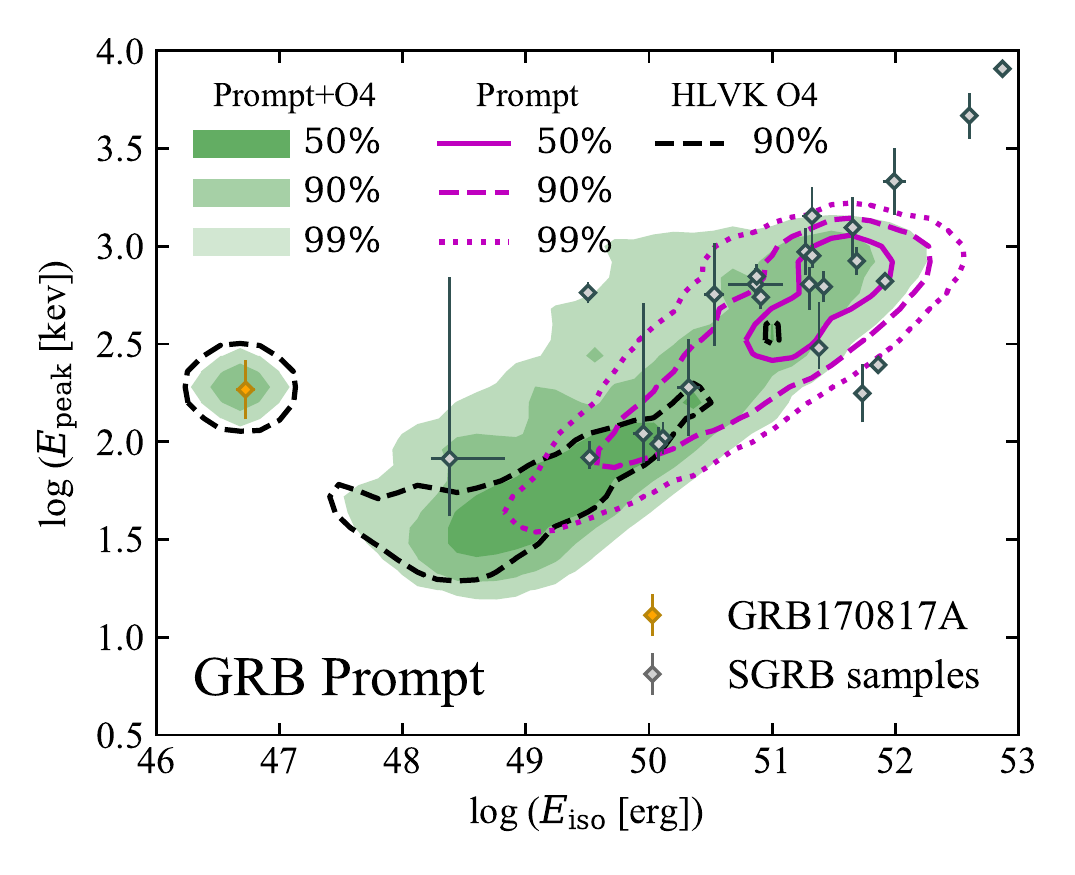}
    \caption{Rest-frame SED peak photon energy $E_\mathrm{peak}$ versus the isotropic-equivalent energy $E_\mathrm{iso}$ for our NSNS population. The filled green colored regions contain $50\%$, $90\%$ and $99\%$ of the binaries both GRB Prompt- and O4-detectable. The magenta lines contain $50\%$, $90\%$ and $99\%$ (solid, dashed and dotted, respectively) of the GRB Prompt-detectable binaries. The black dashed line contains $90\%$ of the O4-detectable binaries. The black dots with error bars represent a SGRB sample for comparison \citep{Salafia2019}. The orange dot is GRB170817A.  }
    \label{fig:amati}
\end{figure}

In Figure \ref{fig:amati} we show the distribution of rest-frame spectral energy distribution (SED) peak energy $E_\mathrm{peak}$ versus the isotropic-equivalent energy $E_\mathrm{iso}$ of events for which both the GW signal and the GRB prompt emission meet our detectability criteria (considering the O4 HLVK network and {\it Fermi}/GBM, green filled contours), and separately those that are detectable in GW (black dashed contours) or by {\it Fermi}/GBM (magenta contours). In particular, different shades in the green regions progressively contain $50\%$, $90\%$ and $99\%$ of joint GRB prompt- and O4-detectable binaries\footnote{The detection rate corresponding to this region is shown by orange lines in Figure \ref{fig:detection} and reported in Table \ref{tab:det_rates}.}. The dashed black line is the $90\%$ confidence region for the O4-detectable binary without the constrain on the GRB prompt detectability. The comparison between the GRB prompt detections by \textit{Fermi}/GBM and the known cosmological population shows a broad consistency with the sample of  short GRBs (SGRBs) with known redshift \citep[][grey diamonds]{Davanzo2018,Salafia2019}. The position of GRB170817A in this plane \citep{Abbott2021_GRB} is shown by the orange diamond, which is consistent by construction with the position of the small island in the left-most part of the plot, which represents events whose emission is dominated by the cocoon shock breakout component.

\section{Discussion and conclusions}

In this article we presented our predictions for the detection rates and properties of KN\ae\  and GRBs (including both prompt and afterglow emission) that will be associated to double neutron star binary mergers to be detected during the next GW detector network run O4, planned to start in December 2022. These predictions are based on a synthetic population of events with an observationally motivated mass distribution and event rate density, for which we computed GW signal-to-noise ratios, KN light curves, GRB afterglow light curves and prompt emission peak photon fluxes, enabling the direct evaluation of the detectability of each emission component for each event in the population. 

KN\ae\  are produced in \frackn\ of mergers in our population, the remaining fraction being massive events that result in a prompt black hole collapse with neither disk nor ejecta (see Figure \ref{fig:mass} in Appendix \ref{apx:mass}). We find light curves that are intrinsically similar to, but on average slightly dimmer than, AT2017gfo (Fig.~\ref{fig:kn}). Despite the larger median distance with respect to events detected in the previous runs, their apparent brightness in most cases (95\% of events with an associated KN) will still exceed the typical limits reached in previous optical counterpart searches, but for a limited time (only the first night in the $g$ band, few nights in the $z$ band), making the detection and identification of these sources more challenging than it had been for AT2017gfo. Our result that most O4 KN\ae\ will be accessible down to current typical EM counterpart search detection thresholds is in line with e.g.\ \citet{Chase_2022}, \citet{Setzer2022} and \citet{SaguesCarracedo2021}.

Relativistic jets are produced in \fracjet\ of the events in our population. Their GRB prompt emission exceeds our assumed limits in only a few percent of the events, with only a minor improvement when considering the deeper thresholds representative of a sub-threshold search in the gamma-ray detector data. A more promising route for the association of a GRB with a NSNS event in O4 will be that of a sub-threshold GW event search in response to a gamma-ray trigger, which results in a joint detection rate of \subGBM\ yr$^{-1}$ in our model, thanks to the expanded GW horizon. 

Radio observations represent the best route for the detection of the relativistic jet afterglow when monitoring a well-localised event. Indeed, radio afterglows are brighter than our `candidate monitoring' detection limits in around one tenth of the simulated events, corresponding to a detection rate of \monzadio\ yr$^{-1}$ (achievable if all candidates are localised to arc-second accuracy through the detection of their KN emission). These predictions indicate that one new relativistic jet counterpart in O4, which would constitute an important new piece of information on these sources, is not unlikely, yet not guaranteed.

For what concerns the observable properties of the relativistic jet counterparts, if a fortunate GRB prompt emission event will be detected, we expect it to be dominated by either the cocoon shock breakout emission component (for events closer than $\sim 100$ Mpc), or more likely by emission from the slower, less energetic material that surrounds the jet core, if a mechanism similar to that which produces the prompt emission of cosmological GRBs extends to two-three times the jet core opening angle (Fig.~\ref{fig:amati}). A due caveat here is that it is unclear to which extent the (poorly known) prompt emission mechanism of GRBs operates outside the jet core and, conversely, the current understanding of shock breakout emission does not extend to highly anisotropic, highly relativistic cases, making any statement on the observable properties of the shock breakout from parts of the cocoon closer to the jet axis highly uncertain. The observable GRB afterglows (Fig.~\ref{fig:after}) are expected to display similar properties as those of GRB170817A, that is, a shallow increase in flux density over a few months after the merger, followed by a peak and a relatively fast decay afterwards. Still, a few percent of the detectable events in our population feature an earlier peak, corresponding to a smaller viewing angle, which would constitute an interesting case study that would bridge the gap between the viewing angles of cosmological GRBs and that of GRB170817A. 

In the last years, several works predicting the joint GW+EM detection rates during O4 have been published or circulated as pre-prints, each focusing on a single or at most two EM counterparts (\citealt{Frostig_2022,Zhu2021, Mochkovitch2021, Saleem2018, Duque2019, Saleem2020, Belgacem2019, Yu2021, Mogushi2019, Howell2019, Abbott2021_GRB}). 
Factoring in the lower local NSNS merger rate density assumed in this work with respect to studies that used the O2 estimate (which was higher by a factor of around three), our joint GW+EM detection rate predictions are in general agreement with most of these previous works. In particular, \citet{Frostig_2022}, \citet{Zhu2021} and \citet{Mochkovitch2021} find a similarly large fraction 60-80\% of KN\ae\ detectable with similar thresholds as ours; factoring in the different fraction of jet-launching events (\fracjet\ in this work, compared to 100\% in the others), our estimate, that up to \monzadiof\ of the afterglows will be detectable in radio, is in agreement with the 20\% estimated by \citet{Duque2019} and \citet{Saleem2018}. 
The prediction that only few percent of the NSNS events detectable in O4 through GW emission will have a detectable short GRB is in line (again factoring in our \fracjet\ fraction of jet-launching systems) with, e.g., \citet{Belgacem2019}, \citet{Howell2019} and \citet{Yu2021}, while \citet{Patricelli_2022}, \citet{Saleem2020} and \citet{Mogushi2019} find somewhat higher fractions (but note that the estimate for sub-threshold GW detections triggered by GRB detections from \citealt{Saleem2020} is in good agreement with ours). 

It is worth stressing the fact that the entirety of these models either assume identical properties for all counterparts, or use empirical parametrizations for the distributions of their properties. The strength of our approach lies in computing the ejecta properties and EM emissions directly from the binary parameters, instead of e.g. extracting EM model parameters randomly from given distributions. 

In this work we worked under the assumption that the GW sky localization areas of O4 NSNS mergers will always be covered down to our assumed representative thresholds by EM counterpart searches. This is clearly not feasible when considering single facilities (due to limitations in the accessible sky and in the time that can be dedicated to each search), but we note that the combined coverage of different facilities can probe very large localization areas, as demonstrated by the searches for EM counterparts of the single-detector event GW190425 (e.g.\ \citealt{Antier2020,Coughlin2019,Lundquist2019,Hosseinzadeh2019}). A more refined assessment of the detection rates that can be realized in practice would require facility-specific simulations of the GW localization and of the actual search strategy (as done e.g.\ in \citealt{Frostig_2022}), which is out of the scope of this work.

As a final remark, our estimates make GRB170817A an extremely lucky event (in line with, e.g., \citealt{Mochkovitch2021}, but see also \citealt{perna2021host}), which is not going to repeat soon. Given the excellent agreement of our model predictions with the short GRB cumulative peak flux distribution observed by {\it Fermi}/GBM and {\it Swift}/BAT (see Figure \ref{fig:lognlogs} in the Appendix), we consider this a robust statement.
Still, we caution that all our predictions are based on loose observational constraints, and carry systematic uncertainties that have not yet been explored, due to the complexity of the full population modeling. The synergy between gravitational and electromagnetic telescopes in future runs will provide us with more observations, allowing to get closer and closer to the real physics of these events. 

\acknowledgments
\textit{Acknowledgments.} We thank the anonymous referee for the valuable comments and suggestions. AC thanks Samuele Ronchini for helpful comments. MC acknowledges funding from the Ministero dell'Università e della Ricerca (MUR) through the Progetti di Rilevante Interesse Nazionale (PRIN) 2017 call, grant number MB8AEZ. OS and GG acknowledge financial support from the MUR, PRIN 2017, grant number 20179ZF5KS.

\acknowledgments
\textit{Data Availability}. The data produced in this work are publicly available on Zenodo at DOI \url{10.5281/zenodo.6900865} through the link \url{https://doi.org/10.5281/zenodo.6900865}. All scripts and files to reproduce the figures in the main text are publicly available through the main author’s Github repository \url{https://github.com/acolombo140/O4NSNS}.

\acknowledgments
\textit{Software.} NumPy \citep{Harris2020}; matplotlib \citep{Hunter2007}; pandas \citep{mckinney2010}; SciPy \citep{Virtanen2020}; Astropy \citep{astropy2013}; PyCBC \citep{pycbc2019}.

\newpage

\appendix

\section{BNS population model}\label{apx:population_model}

\subsection{Mass distribution}\label{apx:mass}
\begin{figure}
    \centering
    \includegraphics[width=0.5\textwidth]{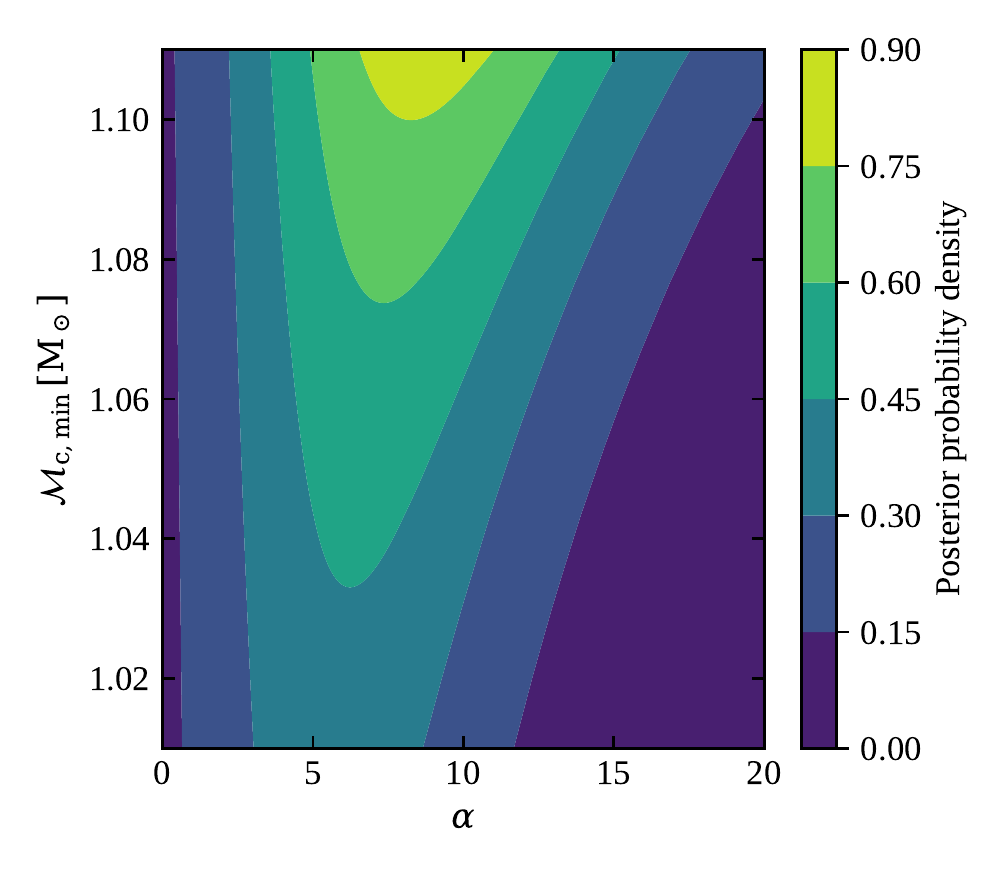}
    \caption{Posterior probability density on our chirp mass probability distribution parameters $\mathcal{M}_{c,\mathrm{min}}$ and $\alpha$. Filled contours show the two-dimensional posterior probability density, with lighter colors corresponding to larger values, as defined in the colorbar.}
    \label{fig:P_mcmin_alpha}
\end{figure}

The mass distribution of merging binary neutron stars is currently not well constrained. Observations of galactic pulsars in double neutron star binaries historically pointed to a narrow distribution centered around $1.33\,\mathrm{M_\odot}$ \citep{Ozel2012,Ozel2016}, but recent studies hint at a bimodal distribution being more likely \citep{Farrow2019}, and the existence of a sub-population detectable with current radio facilities cannot be dismissed \citep[e.g.][]{Pol2019}. Selection effects in GW observations are much simpler, but the only two  detections so far (GW170817, \citealt{GW170817}; and GW190425, \citealt{GW190425}) are insufficient to constrain the shape of the mass distribution. Still, analyses of GW-detected NSs \citep[e.g.][]{Landry2021,population_gwtc3}, combined to GW+pulsar analyses \citep{Galaudage2021}, and arguments based on the incidence of jets \citep{Salafia2022}, clearly point to a relatively broad distribution.  With the aim of defining a simple mass distribution informed only by merging NSNS binaries (as opposed to those obtained by including also the masses of neutron stars in BHNS binaries, \citealt{Landry2021,population_gwtc3}), we devised the following ad-hoc method, which we adopted mainly because of its simplicity, in absence of strong observational constraints. We assumed the component mass probability distribution to be factorized into the chirp mass $\mathcal{M}_c$ probability and the probability of the mass ratio $q=M_2/M_1$ (assumed independent of each other), namely 
\begin{equation}
P(M_1,M_2) = J P(\mathcal{M}_c) P(q)
\end{equation}
where $J(M_1,M_2)=\mathcal{M}_c/M_1^2$ is the Jacobian that relates the two parametrizations \citep{Callister2021}. We then adopted a parametrization for each of these unknown probability distributions, that is
\begin{equation}
    P(\mathcal{M}_c\,|\,\mathcal{M}_{c,\mathrm{min}},\alpha) = \Theta(\mathcal{M}_c-\mathcal{M}_{c,\mathrm{min}})\Theta(\mathcal{M}_{c,\mathrm{max}}-\mathcal{M}_{c})\frac{(1-\alpha)\mathcal{M}_c^{-\alpha}}{\mathcal{M}_{c,\mathrm{max}}^{1-\alpha}-\mathcal{M}_{c,\mathrm{min}}^{1-\alpha}}
\end{equation}
and
\begin{equation}
    P(q\,|\,\beta) = (1+\beta)q^\beta,
\end{equation}
where $\Theta$ is the Heaviside step function and $\alpha$ and $\beta$ are free parameters.
We fixed $\mathcal{M}_{c,\mathrm{max}}=2\,\mathrm{M_\odot}$, which for an equal-mass binary corresponds to $M_1=M_2=2.3\,\mathrm{M_\odot}$ (but we note that this choice does not impact our results significantly). We then looked for maximum-a-posteriori estimates for $\mathcal{M}_{c,\mathrm{min}}$ and $\alpha$ given the observed chirp masses of the two GW-detected events GW170817 ($\mathcal{M}_{c,1} = 1.186\pm 0.001\,\mathrm{M_\odot}$, \citealt{GW170817}) and GW190425 ($\mathcal{M}_{c,2} = 1.44\pm 0.02\,\mathrm{M_\odot}$, \citealt{GW190425}). Following \citealt{Mandel2019}, assuming a simple $\mathcal{M}_{c}^{5/2}$ scaling for the effective searched time-volume and neglecting the small measurement uncertainties, the posterior probability on our two parameters can be written as
\begin{equation}
    P(\mathcal{M}_{c,\mathrm{min}},\alpha\,|\,\mathcal{M}_{c,1},\mathcal{M}_{c,2}) \propto \pi(\mathcal{M}_{c,\mathrm{min}})\pi(\alpha)\frac{P(\mathcal{M}_{c,1}\,|\,\mathcal{M}_{c,\mathrm{min}},\alpha)P(\mathcal{M}_{c,2}\,|\,\mathcal{M}_{c,\mathrm{min}},\alpha)}{\left[\int_{\mathcal{M}_{c,\mathrm{min}}}^{\mathcal{M}_{c,\mathrm{max}}} P(\mathcal{M}_{c}\,|\,\mathcal{M}_{c,\mathrm{min}},\alpha)\mathcal{M}_{c}^{5/2}\,\mathrm{d}\mathcal{M}_{c}\right]^2},
\end{equation}
where $\pi(\mathcal{M}_{c,\mathrm{min}})$ and $\pi(\alpha)$ are the adopted priors. 
Given that the smallest observed chirp mass in a merging Galactic NSNS system is $\sim 1.11\,\mathrm{M_\odot}$ (the chirp mass of J1756-2251, \citealt{Ferdman2014}), we set $\pi(\mathcal{M}_{c,\mathrm{min}})=\Theta(1.11\,\mathrm{M_\odot}-\mathcal{M}_{c,\mathrm{min}})$, while we adopted a broad uniform prior on $\alpha$ in the range $0\leq \alpha \leq 20$. The resulting posterior probability density is shown in Figure~\ref{fig:P_mcmin_alpha},  which shows that the maximum a posteriori probability density is at $\alpha=8.67$ and  $\mathcal{M}_{c,\mathrm{min}}=1.1\,\mathrm{M_\odot}$, the latter being located on the edge of the prior support (which is based on the lowest chirp mass observed in Galactic NSNS binaries).  This tells us that the estimate of $\mathcal{M}_{c,\mathrm{min}}$  is informed by EM Galactic NSNS observations, in addition to GW NSNS merger observations: in that sense, this is a multi-messenger estimate.

In order to constrain the mass ratio distribution parameter $\beta$, we used instead the observed Galactic NSNS mass posteriors from \citealt{Farrow2019}. Their sample comprises $N=10$ NSNSs that will merge within a Hubble time, for each of which they provide $N_s=10^4$ component mass posterior samples. We constructed mass ratio posterior samples $\lbrace q_{i,j}\rbrace_{i=1,...N;\,j=1,..,N_s}$ from these samples, adopting the appropriate mass ordering to ensure $q\leq 1$ for each posterior sample pair. The posterior probability on the $\beta$ parameter based on these samples is then
\begin{equation}
    P(\beta\,|\,\lbrace q_{i,j}\rbrace) \propto \pi(\beta) \prod_{i=1}^{N}\frac{1+\beta}{N_s}\sum_{j=1}^{N_s}q_{i,j}^\beta.
    \label{eq:Pbeta}
\end{equation}
We adopted a uniform prior $\pi(\beta)$ in the range $0\leq \beta \leq 40$. 
\begin{figure}
    \centering
    \includegraphics[width=0.5\textwidth]{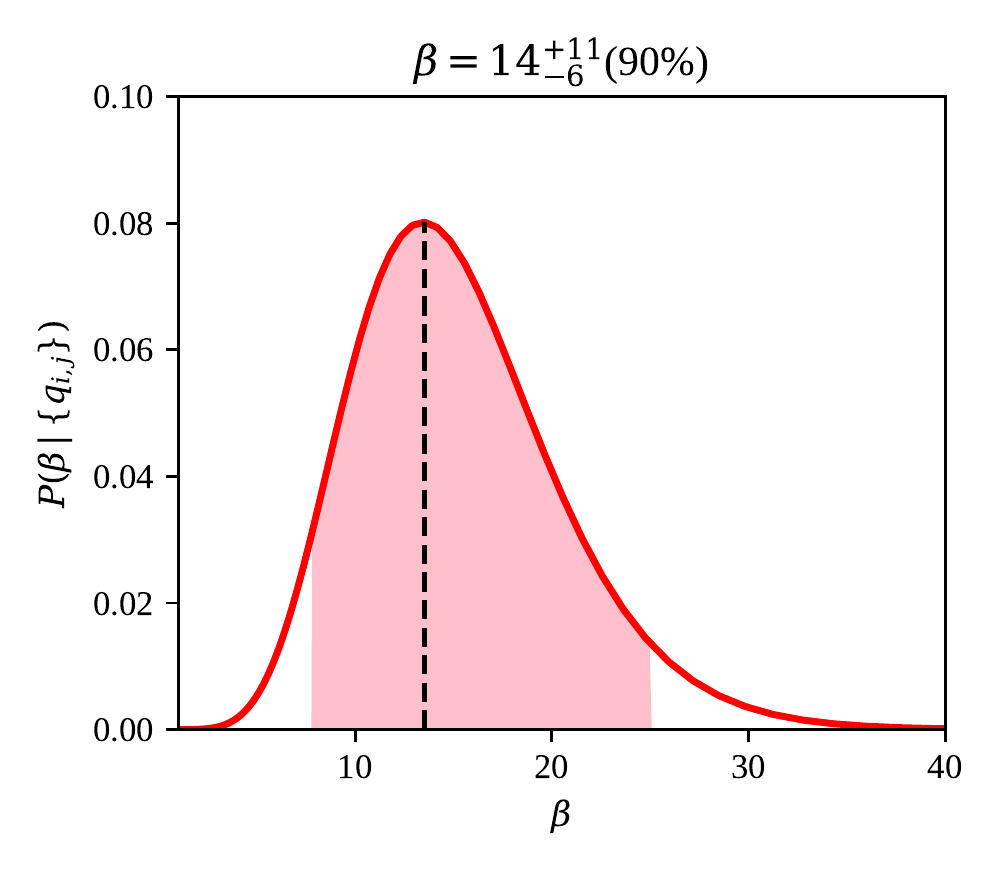}
    \caption{Posterior probability density on the $\beta$ parameter of our assumed mass ratio probability distribution parametrization. The red solid curve shows the result of Eq.~\ref{eq:Pbeta} using the mass information on 10 Galactic NSNS binaries that merge within a Hubble time from \citet{Farrow2019}. The pink shaded area shows the 90\% credible interval, while the vertical dashed line marks the maximum \textit{a posteriori}.}
    \label{fig:beta_posterior}
\end{figure}
The resulting posterior probability distribution is shown in Figure~\ref{fig:beta_posterior}, which shows a large uncertainty, but a well-defined peak at $\beta=14$. 
\begin{figure}
    \centering
    \includegraphics[width=0.5\textwidth]{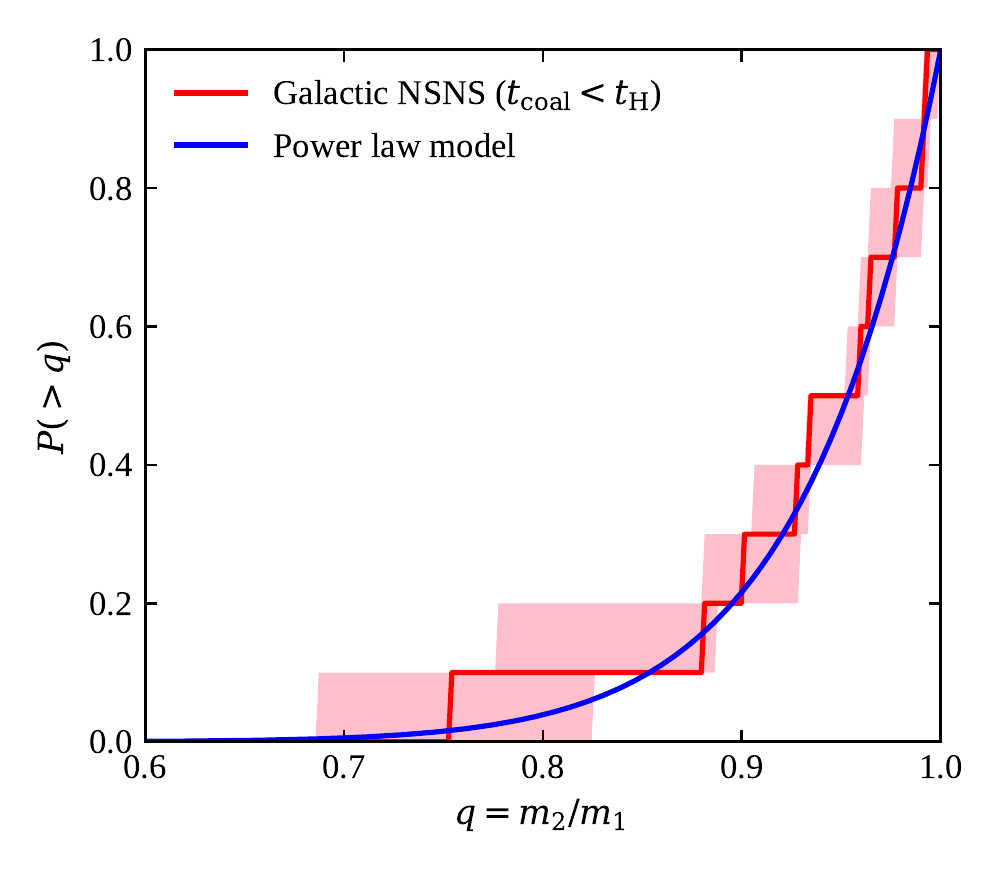}
    \caption{Cumulative mass ratio probability density in our mass distribution model (blue solid line), compared to the observed Galactic NSNS (with coalescence time less than the Hubble time) mass ratio cumulative distribution (red line: median; pink area: 90\% uncertainty region -- based on the data from \citealt{Farrow2019}).}
    \label{fig:q_cumulative}
\end{figure}
Figure~\ref{fig:q_cumulative} compares the observed Galactic NSNS mass ratio cumulative distribution and our mass ratio distribution model $P(q\,|\,\beta)$ with the maximum-a-posteriori value $\beta=14$. 
\begin{figure}
    \centering
    \includegraphics[width=0.6\textwidth]{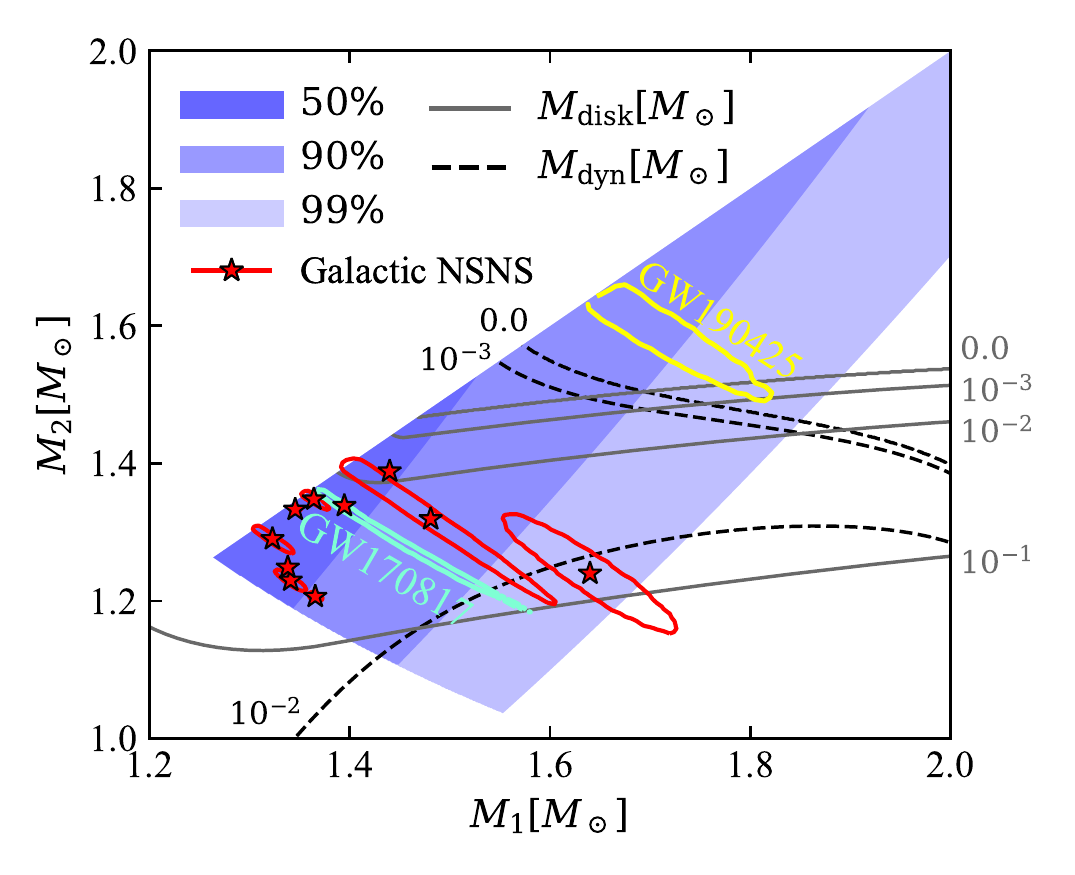}
    \caption{$M_1, M_2$ plane showing the mass distribution for our NSNS population. The filled blue colored regions contain $50\%$, $90\%$ and $99\%$ of the binaries. The black dashed lines and the grey lines represent respectively the contours for the predicted dynamical ejecta and disk mass, assuming the SFHo EoS. Red stars and contours show the best fit and $90\%$ credible regions for the known Galactic NSNS \citep{Ozel2016,Farrow2019} systems that merge within a Hubble time. Yellow and aquamarine lines represent the $50\%$ confidence regions for the component masses in GW190425 \citep{Abbott2020_GW190425} and GW170817 \citep{Abbott2019_GW170817_properties}, both constructed using the publicly available low-spin-prior posterior samples. }
    \label{fig:mass}
\end{figure}
Figure~\ref{fig:mass} compares the resulting $P(M_1,M_2\,|\,\mathcal{M}_{c,\mathrm{min}},\alpha,\beta)$ mass distribution model with the observations on the $(M_1,M_2)$ plane. It also shows iso-contours of ejecta and accretion disk mass obtained with our adopted fitting formulae \citep{krouger2020,Barbieri2019_2} and equation of state (EoS) - SFHo -, which helps in visualizing the absence of EM counterparts for events in the upper right corner of the plane, and the general trends in the distribution of ejecta and disk masses in the population. 
\begin{figure}
    \centering
    \includegraphics[width=\textwidth]{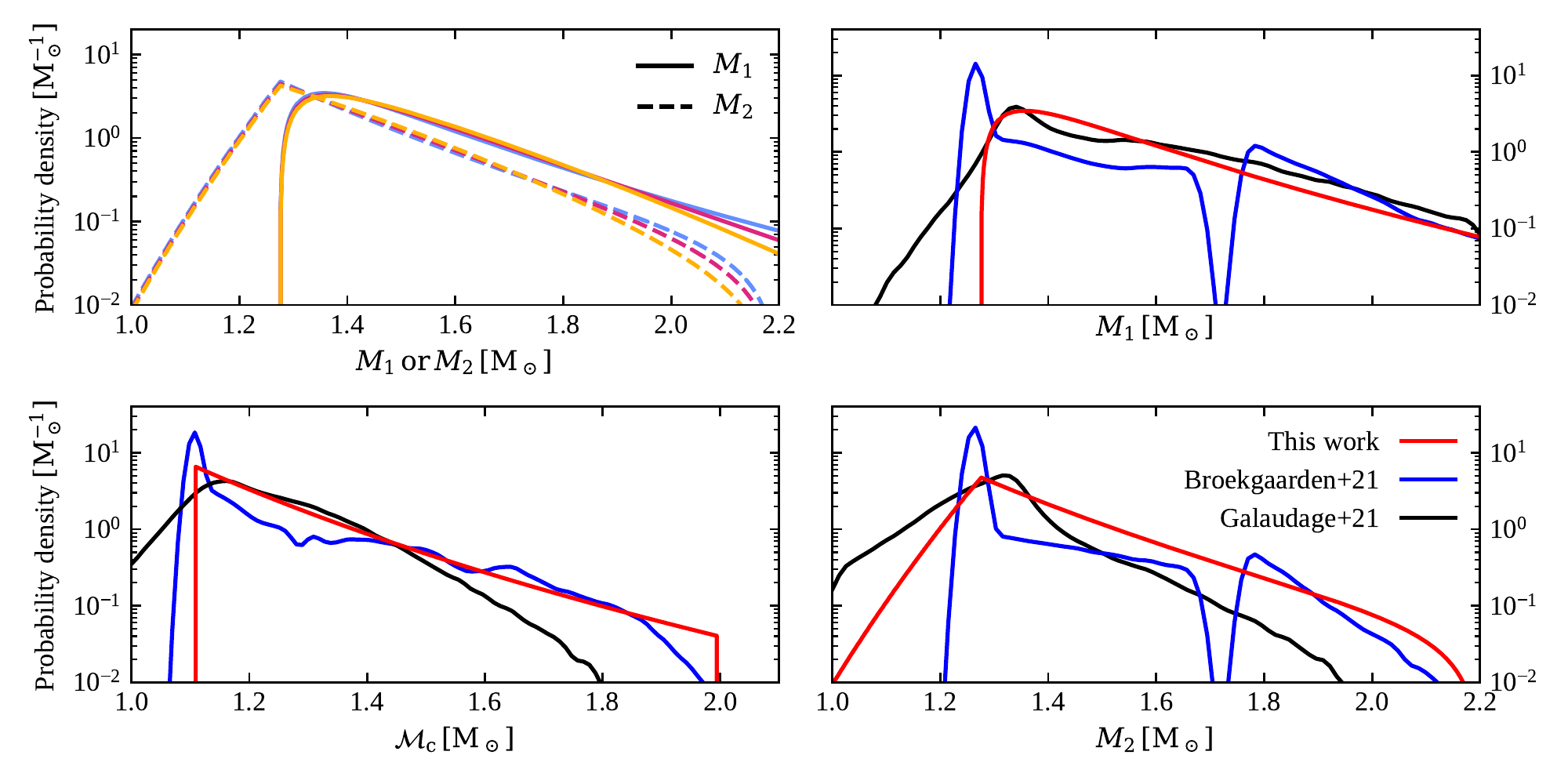}
    \caption{Component mass probability distribution comparisons. Upper-left panel: NSNS component mass probability distributions (solid lines: primary mass; dashed lines: secondary mass) from our model, assuming three different parametrizations of the chirp mass probability distribution, namely a power law (purple -- the fiducial model described in the text), a decreasing exponential  (orange) and a Gaussian tail (light blue). Other panels: comparison of our component mass (top-right: primary; bottom-right: secondary) and chirp mass (lower-left) probability distributions (red lines) with those from a state-of-the-art population synthesis model \citep[the fiducial model from][blue lines]{Broekgaarden2021} and of an observational study that combines Galactic and GW NSNS measurements \citep[][black lines, obtained considering their median distributions]{Galaudage2021}.}
    \label{fig:mass_distrib_comparisons}
\end{figure}
We note that changing the chirp mass parametrization to either an exponential $P(\mathcal{M}_c)\propto \Theta(\mathcal{M}_c-\mathcal{M}_{c,\mathrm{min}})\exp(-\mathcal{M}_c/\mathcal{M}_{c,\mathrm{scale}})$ or a Gaussian tail $P(\mathcal{M}_c)\propto \Theta(\mathcal{M}_c-\mathcal{M}_{c,\mathrm{min}})\exp\left[-\left(\mathcal{M}_c/\mathcal{M}_{c,\mathrm{scale}}\right)^2\right]$ does not alter significantly our results, as demonstrated in Figure~\ref{fig:mass_distrib_comparisons} (left-hand panel).

It is instructive to compare our mass probability distribution with others in the literature. To that purpose, we show in the right-hand panels of Figure~\ref{fig:mass_distrib_comparisons} a comparison of the probability distributions of component and chirp masses implied by our result (red lines) with the corresponding distributions from a recently published population synthesis model \citep[][their fiducial model]{Broekgaarden2021} based on the COMPAS code \citep{Riley2022}, and with the result of the study by \citealt{Galaudage2021}, which models the Galactic NSNS population and the GW-detected NSNS binaries together. These comparisons show that, despite the large uncertainties and the simplifying assumptions, our results fall in a reasonably similar range as other results based on more refined methodologies. Last, but not least, our mass distribution combined with our choice of the EoS leads to a large fraction of remnants that satisfy the basic requirements for the launch of a relativistic jet by the \citealt{Blandford1977} process, namely a hyper-massive NS or a BH remnant and a non-negligible accretion disk, as required by the high observed incidence of jets \citep[see][who discuss this argument and the implied mass distribution constraints in detail]{Salafia2022}.

\subsection{Redshift distribution}\label{apx:redshift}

Merging binary neutron stars are thought to form either from isolated stellar binaries or in dense stellar environments such as stellar clusters, in which dynamical interactions can play a non-negligible role in their formation and evolution \citep{Smarr1976,Srinivasan1989,PortegiesZwart1998,Bhattacharya1991}.  Taking into account the strong dependence  of the GW-driven coalescence timescale  $t_\mathrm{c,GW}$ on the binary separation $a$, $t_\mathrm{c,GW}\propto a^4$, and expressing the  probability distribution of $a$ as a power law with index $x$, namely $\mathrm{d}P/\mathrm{d}a\propto a^x$, the probability distribution of the delay time between the start of the GW-driven inspiral and the coalescence is  $\mathrm{d}P/\mathrm{d}t_\mathrm{c,GW}=(\mathrm{d}P/\mathrm{d}a)( \mathrm{d}a/\mathrm{d}t_\mathrm{c,GW})\propto  t_\mathrm{c,GW}^{-3/4+x/4}$ \citep{Piran1992}. Being the result of a diverse and complex range of processes, it is reasonable to expect the separation distribution $\mathrm{d}P/\mathrm{d}a$ to be close to uniform in the logarithm, and hence $x\sim -1.$  This translates into a delay time distribution $\mathrm{d}P/\mathrm{d}t_\mathrm{c,GW}$ that is also close to uniform-in-log, and the $x/4$ dependence ensures that this remains approximately true unless $x$ is very large in absolute value. When the coalescence timescale $t_\mathrm{c,GW}$ is longer than $t_\mathrm{SN2}$, the time elapsed between the birth of the binary and the formation of the second neutron star, then the delay $t_\mathrm{d}$ between the binary formation and its coalescence also follows the same power law; conversely, for very short GW coalescence timescales, the delay time $t_\mathrm{d}\sim t_\mathrm{SN2}$. These arguments lead to a delay time distribution of the form
\begin{equation}
    \frac{\mathrm{d}P}{\mathrm{d}t_\mathrm{d}}\propto \left\lbrace\begin{array}{cc}
         t_\mathrm{d}^{-1} & t_\mathrm{d}\geq \langle t_\mathrm{SN2} \rangle \\
         0 & t_\mathrm{d}<  \langle t_\mathrm{SN2} \rangle
    \end{array}\right.,
\end{equation}
where $\langle t_\mathrm{SN2}\rangle$ is the mean time to the second supernova, which we take as $\langle t_\mathrm{SN2}\rangle=50\,\mathrm{Myr}$, appropriate for the lightest neutron star progenitors. This distribution is broadly consistent with the results of detailed binary stellar population synthesis models \citep[e.g.][]{Dominik2012}. With the further simplifying assumption of a constant fraction of stellar mass going into binaries that end up as double neutron stars throughout the history of the Universe, the cosmic NSNS merger rate density can be then modelled as
\begin{equation}
    \dot\rho(t)=\frac{\mathrm{d}^2 N}{\mathrm{d}V\mathrm{d}t} \propto \int_t^\infty \dot \rho_\star(t') \frac{\mathrm{d}P}{\mathrm{d}t_\mathrm{d}}(t'-t)\mathrm{d}t',
\end{equation}
where $t=t(z)$ is the lookback time corresponding to redshift $z$, $\mathrm{d}V$ is the comoving volume element, and $\dot \rho_\star$ is the cosmic star formation rate density, for which we adopt the analytical form given in \citealt{Madau2014}. 

\subsection{Local rate density}\label{apx:R0}

The normalization of the assumed NSNS merger rate density, namely the local neutron star merger rate density $\dot\rho(0)=R_0$, was set based on self-consistency of the total number of NSNS detections in the three past observing runs of the advanced GW detector network and the number expected given our chosen mass and redshift distributions. To do this in practice, we needed to estimate the effective time-volume searched by the LIGO-Virgo network during the three observing runs O1, O2 and O3, which can be defined as \citep[e.g.][]{Tiwari2018}
\begin{equation}
 V_\mathrm{eff}=f_\mathrm{det}(<z_\mathrm{max})\int_0^{z_\mathrm{max}} \frac{\dot\rho(z)}{\dot\rho(0)}\frac{\mathrm{d}V}{\mathrm{d}z}\frac{\mathrm{d}z}{1+z},
\end{equation}
where 
$\mathrm{d}V/\mathrm{d}z$ is the differential comoving volume \citep{Hogg1999}, $z_\mathrm{max}$ is any redshift beyond the O3 GW detectability horizon, and $f_\mathrm{det}(<z_\mathrm{max})$ is the fraction of detectable NSNS mergers within $z_\mathrm{max}$. To estimate the latter, we took the publicly available LVK Collaboration O1+O2+O3 sensitivity study Monte Carlo samples \citep{gwosc_sensitivity_study}, we re-sampled them to reflect our assumed mass and redshift distributions, and then computed $f_\mathrm{det}(<z_\mathrm{max})$ as the fraction of events that satisfied our detectability cut $\mathrm{SNR_{net}}\geq 12$ over the total within $z_\mathrm{max}$. This resulted in $V_\mathrm{eff}=5.1\times 10^{-3}\,\mathrm{Gpc^{3}}$. Given the  actual number $N_\mathrm{obs}=2$ of observed NSNS events that satisfy the same cut (i.e.\ GW170817 and GW190425), and given the total O1+O2+O3 effective observing time $T=1.23\,\mathrm{yr}$ \citep[][representing the total time span of observing periods with at least one active detector]{GWTC3}, we obtained the posterior on the local merger rate density $R_0$ (conditional on our assumed mass and redshift distribution)
\begin{equation}
 P(R_0\,|\,N_\mathrm{obs})\propto R_0^{N_\mathrm{obs}}\exp\left(-R_0 V_\mathrm{eff}T\right)\pi(R_0),
\end{equation}
where we adopted the Jeffreys prior $\pi(R_0)=R_0^{-1/2}$. The resulting median and symmetric 90\% credible interval are $R_0=\RoNSNS$, which therefore includes the statistical Poisson uncertainty stemming from the small number of observed events, but not any model systematic uncertainty (which would result in a larger uncertainty, probably more akin to the ones from \citealt{population_gwtc3}), which is not explored here.

\section{EM emission models}\label{apx:em_model}
In the following section, we briefly describe the models employed to compute the EM emissions from the NSNS mergers in our synthetic population. We refer to \citealt{Perego2017}, \citealt{Barbieri2019}, \citealt{Barbieri2020}, \citealt{Breschi2021}, \citealt{Salafia2019} and \citealt{Salafia2019b} for more detailed descriptions. 

\subsection{Ejecta}
\begin{figure}
    \centering
    \includegraphics[width=0.6\textwidth]{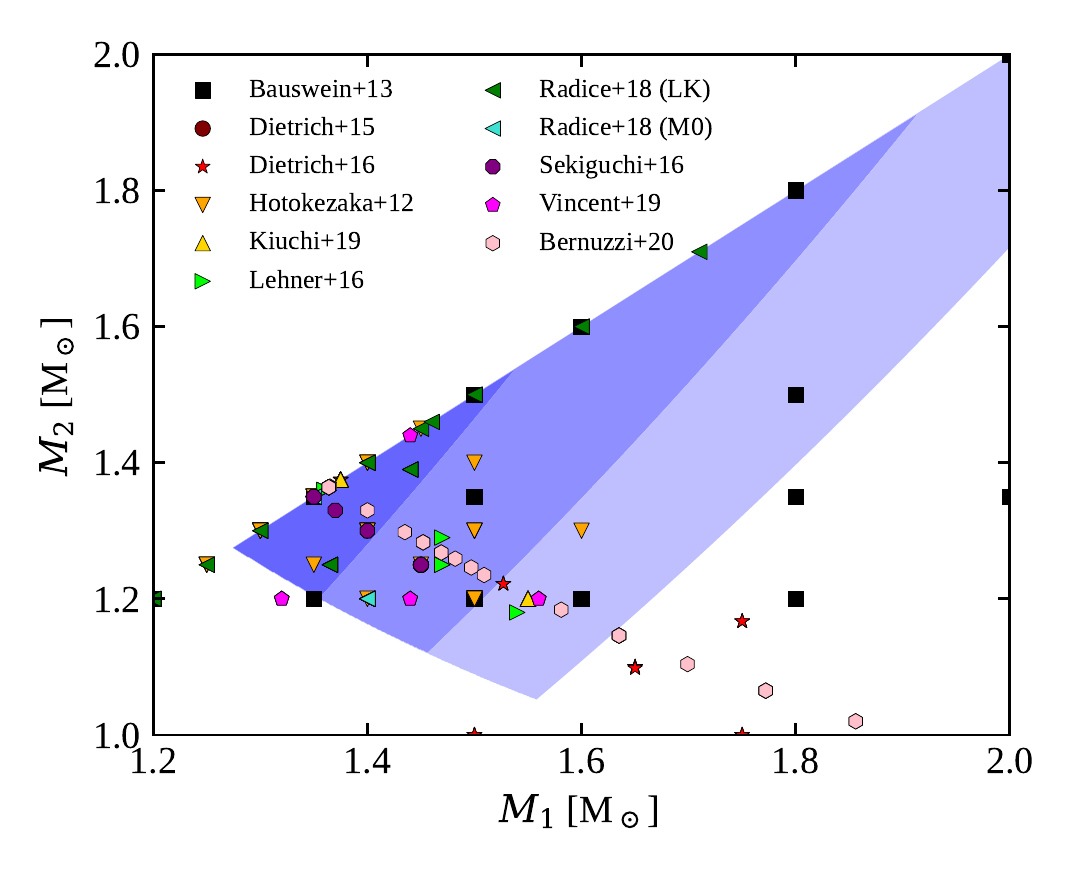}
    \caption{Numerical-relativity simulations used to calibrate the ejecta and disk mass fitting formulae compared to our assumed mass distribution on the $M_1, M_2$ plane. Filled contours show the smallest areas containing 50\%, 90\% and 99\% of the binaries in our NSNS population (same color coding as in Figure \ref{fig:mass}). Colored symbols mark the pairs of neutron star gravitational masses corresponding to the simulations in the calibration set. Different symbols are used for simulations described in different references. The ejecta mass fitting formula \citep{krouger2020} is calibrated on the data from \citealt{Hotokezaka2013}, \citealt{Bauswein2013},  \citealt{Dietrich2015}, \citealt{Lehner2016}, \citealt{Sekiguchi2016}, \citealt{Dietrich2017} and \citealt{Kiuchi2019}; the ejecta velocity fitting formula is calibrated on data from \citealt{Radice2018_2} (using only the simulations with neutrino leakage -- LK); the disk mass fitting formula \citep{Barbieri2020_2} is calibrated on data from \citealt{Radice2018_2} (both simulations with neutrino leakage -- LK -- and those with M0 transport -- M0), \citealt{Vincent2020}, \citealt{Kiuchi2019} and \citealt{Bernuzzi2020}.}
    \label{fig:mass_support}
\end{figure}

We divide the material ejected in a NSNS merger in three broad classes. 
The first component, the \textit{dynamical} ejecta, is material  ejected on dynamical timescales ($\sim$ms) by either tidal forces operating during the last phases of the inspiral (which launches cold, highly neutron-rich material mainly close to the equatorial plane), or by shocks generated in the collision of the neutron star cores (which generates a higher-entropy, less neutron-rich component that is launched more isotropically). Depending on the specific angular momentum distribution of the NS decompressed matter, a certain fraction can be centrifugally supported, forming  an accretion disk around the merger remnant.

The accretion disk can then produce additional ejecta in the form of winds, on longer timescales. We divide these into `wind' ejecta, carried along directions close to the polar axis by the neutrino flux produced in the inner, hotter regions of the disk during the neutrino-dominated phase (typically lasting few tens of ms, e.g.\ \citealt{just2015}), and  `secular' ejecta, released due to viscous angular momentum transport on the viscous time scale (of the order of $1\,\mathrm{s}$, e.g.\ \citealt{just2015}) with a fairly isotropic distribution. 
In our model, we compute the ejecta properties, as a function of the binary parameters (namely the component masses and the EoS), using fitting formulae based on numerical simulations of the merger and post-merger dynamics. In particular we adopt the fitting formulae from \citealt{krouger2020} and \citealt{Radice2018_2}, in order to compute the mass and average velocity of the dynamical ejecta, respectively. We instead compute the accretion disk mass using the fitting formula from \citealt{Barbieri2020}, whose predictions are consistent with both symmetric and asymmetric NSNS merger numerical simulations presented in \citealt{Radice2018_2}, \citealt{Kiuchi2019}, \citealt{Bernuzzi2020} and \citealt{Vincent2020}. In Figure \ref{fig:mass_support} we compare our mass distribution with the masses of the neutron stars in the simulations used to calibrate the fitting formulae for the ejecta properties. The comparison shows that most of our neutron star binaries have masses in ranges over which the fitting formulae have a fair number of calibration points. Still, we caution that for a $\lesssim 10\%$ fraction of the binaries (high-mass, asymmetric systems)  the disk and ejecta properties that we obtain from the fitting formulae are essentially extrapolations. Finally, we compute the masses of the wind and secular ejecta by assuming that fixed fractions of these ejecta $\xi_\mathrm{w}=0.05$ and $\xi_\mathrm{s}=0.2$, respectively, go into these components \citep{Perego2014,Fernandez2016,Siegel2017,Fujibayashi2018}. For each ejecta class we finally assume angular profiles of rest-mass density, average velocity and opacity identical to model ANI-DVN from \citealt{Breschi2021}.

\subsection{Kilonova}\label{apx:kn}
\begin{figure}
    \centering
    \includegraphics[width=1\textwidth]{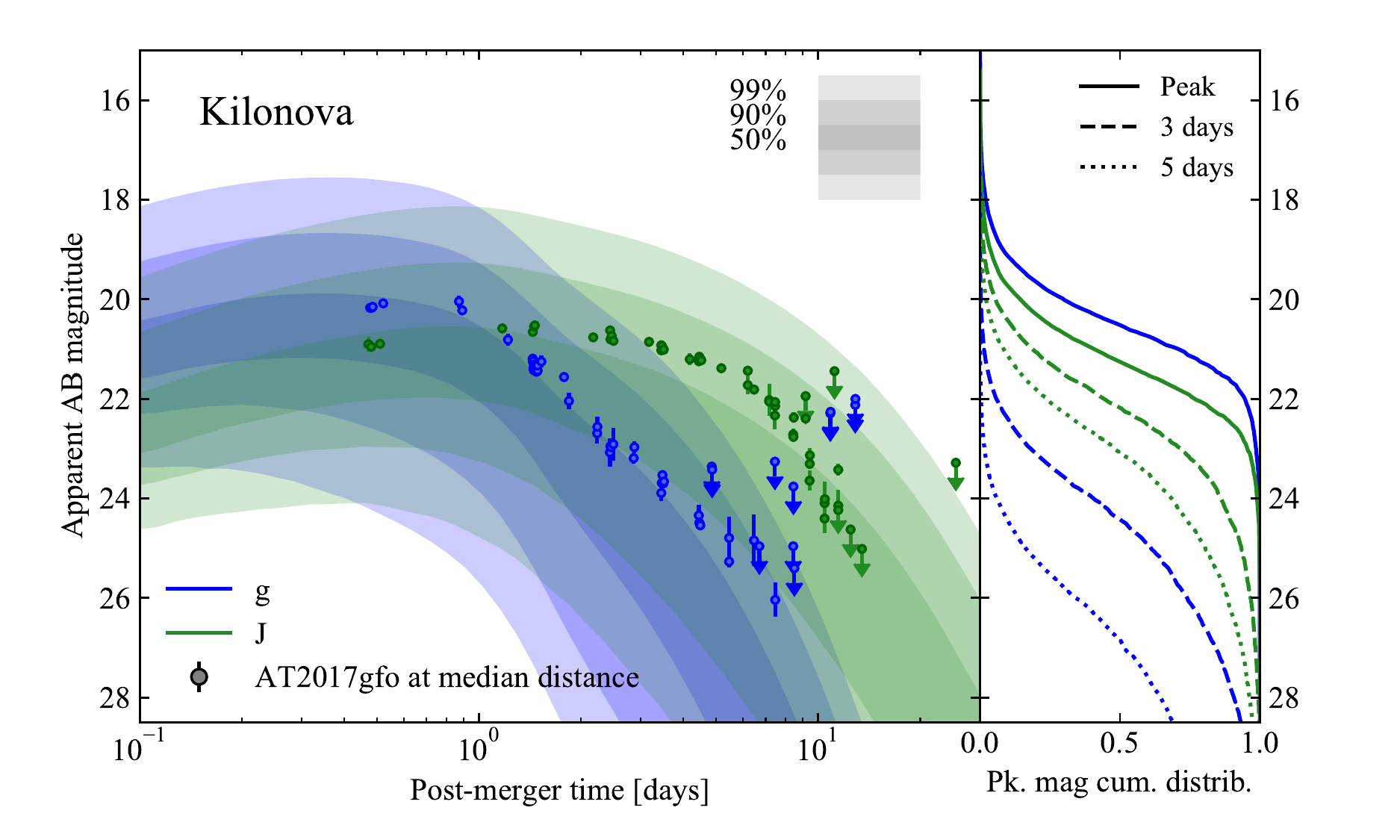}
    \caption{Same as Figure \ref{fig:kn}, but for the g band (484 nm, blue) and for the J band (1250 nm, green).}
    \label{fig:kn_gJ}
\end{figure}

We compute the kilonova light curves following \citealt{Perego2017} (based in part on the works by \citealt{Grossman2014} and \citealt{Martin2015}), with the additions described in \citealt{Barbieri2019} and \citealt{Breschi2021}. In brief, the computation is based on a semi-analytical model in which axisymmetry relative to the direction of the binary angular momentum is imposed. The ejecta, assumed to be in homologous expansion, are divided into polar angle bins, and thermal emission at the photosphere of each angular bin along radial rays is computed following \citealt{Grossman2014} and \citealt{Martin2015}, taking into account the projection of the photosphere in each bin. 

Figure \ref{fig:kn_gJ} shows the time evolution of the distribution of KN brightness in the $g$ and $J$ bands for our population, computed with the above model and the prescriptions described in the text to link the KN ejecta properties to those of the progenitors, similarly as in Figure \ref{fig:kn} (which referred to the $g$ and $z$ band, instead).

\subsection{Relativistic jet}\label{apx:rel_jet}


We assume the relativistic jet to be launched by the Blandford-Znajek mechanism, which requires a spinning BH surrounded by a magnetized accretion disk (\citealt{Blandford1977} and \citealt{Komissarov2001}). In the context of NSNS mergers, in order for these requirements to be fulfilled, the remnant must collapse to a BH on a time shorter than the disk viscous timescale, which restricts the possible merger outcomes to hypermassive neutron stars and prompt BH collapse only, that is, to remnants with a mass $M_\mathrm{rem}\geq 1.2\,M_\mathrm{TOV}$ \citep[e.g.][]{Salafia2022}, with the additional requirement that an accretion disk must form, which in prompt collapsing cases is possible if the binary is asymmetric \citep{Bernuzzi2020}. When these conditions are fulfilled, we compute the jet total injected energy as in \citealt{Barbieri2019}
\begin{equation}
    E_\mathrm{jet,0}=\epsilon(1-\xi_\mathrm{w}-\xi_\mathrm{s})M_\mathrm{disk}c^2\Omega_\mathrm{H}^2 f(\Omega_\mathrm{H}),
\end{equation}
as a function of the disk mass $M_\mathrm{disk}$ and the remnant BH spin $\chi_\mathrm{BH}$ via the quantities $\Omega_\mathrm{H}=\chi_\mathrm{BH}/2(1+\sqrt{1-\chi_\mathrm{BH}^2})$ and $f(\Omega_\mathrm{H})=1+1.38\Omega_\mathrm{H}^2-9.2\Omega_\mathrm{H}^4$ \citep{Tchekhovskoy2010}. The dimensionless constant $\epsilon$ is fixed by imposing the accretion-to-jet energy conversion efficiency $\eta=\epsilon\,\Omega_\mathrm{H}^2 f(\Omega_\mathrm{H})$ to be $\eta=10^{-3}$ when $\chi_\mathrm{BH}=0.71$, therefore matching the inferred efficiency in GW170817 \citep{SalafiaGiacomazzo2021}. This leads to $\epsilon=0.022$.

Part of this energy is spent by the jet in its propagation through the ejecta cloud. Following \citealt{Duffell2018}, we assume the energy needed for the jet to successfully break out of the ejecta to be $E_\mathrm{bkt}=0.05 \theta_\mathrm{j,0}^2 E_\mathrm{ej}$, where we set the jet opening angle at launch $\theta_\mathrm{j,0}=15^\circ$ and we compute the ejecta energy as the sum of the isotropic-equivalent energies (averaged within an angle $\theta_\mathrm{j,0}$ from the polar axis, and accounting for their assumed angular profiles -- see the previous section and \citealt{Breschi2021}) of the three considered ejecta components, $E_\mathrm{ej}=E_\mathrm{iso,ej,dyn}+E_\mathrm{iso,ej,wind}+E_\mathrm{iso,ej,sec}$.
If $E_\mathrm{jet,0}\leq E_\mathrm{bkt}$, we consider the jet to be choked during the propagation and we neglect its emission (this happens in \fracbkt\ of jet-launching systems in our population); otherwise, we assume the jet to successfully break out, with an available energy $E_\mathrm{jet}=E_\mathrm{jet,0}-E_\mathrm{bkt}$.

We assume jets that successfully break out to be endowed with a jet structure (angular energy and bulk Lorentz factor profiles) featuring a uniform core of half-opening angle $\theta_\mathrm{c}$, surrounded by ``wings'' with power-law decreasing energy density and Lorentz factor. Explicitly 
\begin{align}
    &\frac{dE}{d\Omega}(\theta)=\frac{\epsilon_\mathrm{c}}{1+(\theta/\theta_\mathrm{c})^{s_\mathrm{E}}},\\
    &\Gamma(\theta)=1+\frac{\Gamma_\mathrm{c}-1}{1+(\theta/\theta_\mathrm{c})^{s_\Gamma}},
\end{align}
where $\epsilon_\mathrm{c}=(E_\mathrm{jet}-E_\mathrm{bkt})/\pi\theta_\mathrm{c}^2$ is the core energy per unit solid angle and $\Gamma_\mathrm{c}$ is the core Lorentz factor. We keep the structure parameters identical across the population, fixing $\theta_{c}=3.4^\circ$, $s_\mathrm{E}=5.5$, $\Gamma_\mathrm{c}=251$ and $s_\mathrm{\Gamma}=3.5$, which are the best-fit values for the GRB170817A afterglow from \citealt{Ghirlanda2019}. 

\begin{figure}
    \centering
    \includegraphics[width=0.6\textwidth]{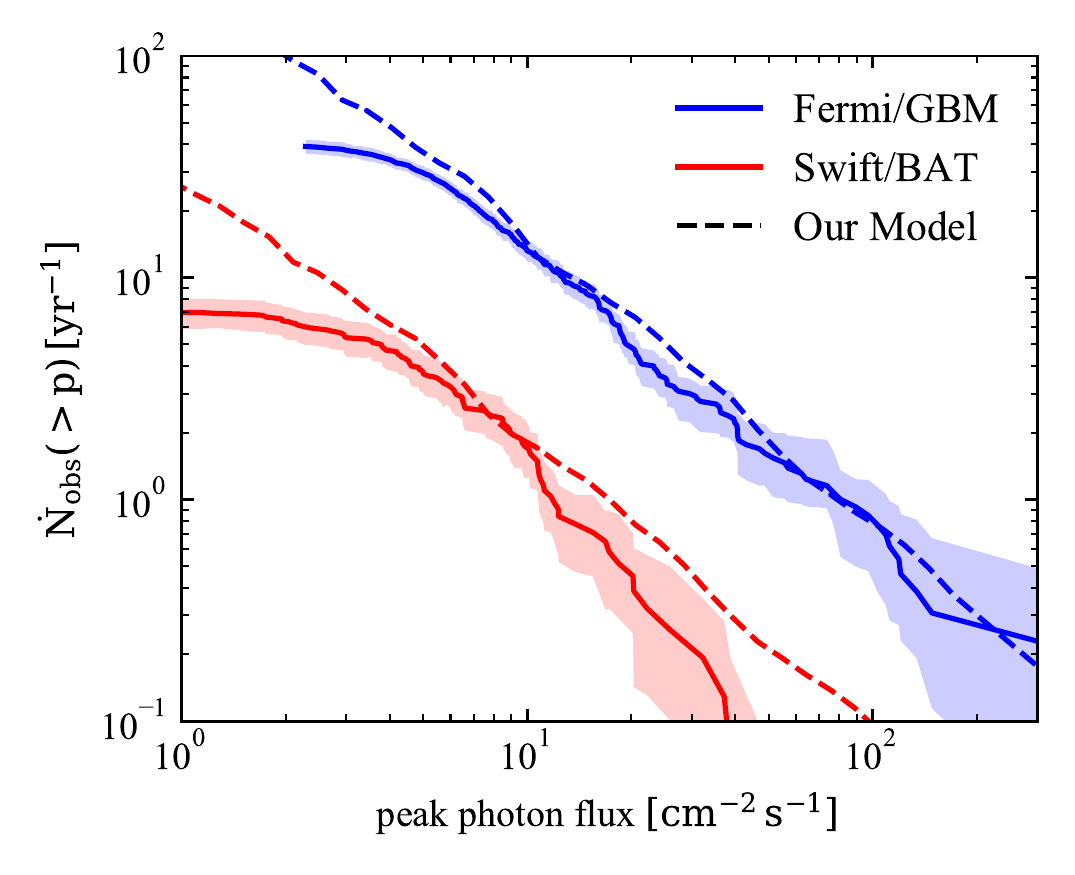}
    \caption{{\it Fermi}/GBM ({\it Swift}/BAT) observed inverse cumulative distribution of 64ms (20ms)-binned photon fluxes in the 10--1000 (15--150) keV band (the colored band shows the $90\%$ confidence band due to Poisson and measurement uncertainties) compared with our model (dashed line).}
    \label{fig:lognlogs}
\end{figure}
\subsubsection{Gamma-ray burst prompt emission}\label{apx:prompt}

As stated in the main text, we compute the prompt emission spectrum following \citealt{Salafia2015, Salafia2019}, assuming the conversion efficiency of jet energy into radiation to be $\eta_\gamma=0.15$ in regions of the jet with $\Gamma\geq 10$, and zero otherwise.  The isotropic equivalent specific luminosity at observer frequency $\nu$, as measured by an observer who sees the jet under a viewing angle $\theta_\mathrm{v}$, and under the assumption of a viewing-angle-independent emission duration $T$, is then given by \citep{Salafia2015}
\begin{equation}
L_\mathrm{\nu,iso}(\nu,\theta_v)= \frac{\eta_\gamma}{T} \int_{0}^{\theta_\gamma}\sin\theta\,\mathrm{d}\theta\int_0^{2\pi}\mathrm{d}\phi \,S_{\nu'}(\nu(1+z)/\delta)\frac{\delta^2}{\Gamma}\frac{\mathrm{d}E}{\mathrm{d}\Omega},
\end{equation}
where $z$ is the source redshift, $\theta_\gamma$ is the angle such that $\Gamma(\theta_\gamma)=10$ (which is $\theta_\gamma=8.7^\circ$ with our parameters), $\delta$ is the relativistic Doppler factor of material located at spherical angular coordinates $(\theta,\phi)$, and $S_{\nu'}$ is the comoving spectral shape, which we assume to be a power law with an exponential cut-off, $S_\mathrm{\nu'}\propto (\nu')^a \exp\left[-(1+a)\nu'/\nu'_\mathrm{p}\right]$, with $a=0.24$ and $h\nu'_\mathrm{p}=3\,\mathrm{keV}$ ($h$ here is Planck's constant), similarly as in \citealt{Salafia2019}, and the normalization is such that $\int S_{\nu'}d\nu' = 1$.

To account for the contribution of a putative shock breakout emission component, for viewing angles $\theta_\mathrm{v}<60^\circ$ we include an additional emission component with identical properties as GRB170817A, namely an isotropic-equivalent luminosity $L_\mathrm{iso}=10^{47}\,\mathrm{erg/s}$ and a cut-off power law spectrum (same shape as the assumed prompt emission spectrum) with $h\nu_\mathrm{p}=E_\mathrm{peak}=185\,\mathrm{keV}$ and $a=0.38$ \citep{Abbott2021_GRB}.

From the specific luminosity we obtain the photon flux in the $[h\nu_\mathrm{0}, h\nu_\mathrm{1}]$ observing band as \begin{equation}
p_{[h\nu_\mathrm{0},h\nu_\mathrm{1}]}=\frac{1}{4\pi d_\mathrm{L}^2}\int_{\nu_\mathrm{0}}^{\nu_\mathrm{1}}\frac{L_\mathrm{\nu,iso}(\nu(1+z))}{h\nu(1+z)}\mathrm{d}\nu,
\end{equation}
where $d_L$ is the source luminosity distance. Figure \ref{fig:lognlogs} shows the inverse cumulative distributions of photon fluxes in the [10,1000] keV (blue) and [15,150] keV (red) bands for our population (dashed lines), and the corresponding distributions for short GRBs observed by {\it Fermi}/GBM and {\it Swift}/BAT, respectively (solid lines, with the shaded area showing the 90\% confidence regions, including both measurement uncertainties and Poisson count statistics). The distributions for our model are computed accounting for the duty cycle and field of view factors for each instrument, for a fair comparison. 

\subsubsection{Afterglow}\label{apx:aft}
The afterglow emission model is described in \citealt{Salafia2019} and \citealt{Barbieri2019}. In brief, this is a semi-analytical model based on standard afterglow theory \citep{Sari1998,Panaitescu2000}, extended to the case of an inhomogeneous jet and an off-axis viewing angle. The shock dynamics model is valid in both the ultra-relativistic and the non-relativistic regime, but it does not include lateral expansion. The emission model only includes synchrotron emission, assuming constant (throughout the evolution and independent of the angle) the relativistic electron energy `equipartition' parameter $\epsilon_\mathrm{e}=0.1$, the magnetic field equipartition parameter $\epsilon_\mathrm{B}=10^{-3.9}$ and   $p=2.15$, based on GRB170817A \citep{Ghirlanda2019}. Syncrotron self-absorption is included. The interstellar medium in which the shock expands is assumed to have a uniform number density  of $5\times 10^{-3}\,\mathrm{cm^{-3}}$. The surface brightness is computed locally based on the fluid properties behind the shock, and the flux is computed by integrating the surface brightness over equal-arrival-time surfaces at the relevant viewing angle, accounting for relativistic effects.

\begin{figure}
    \centering
    \includegraphics[width=0.6\textwidth]{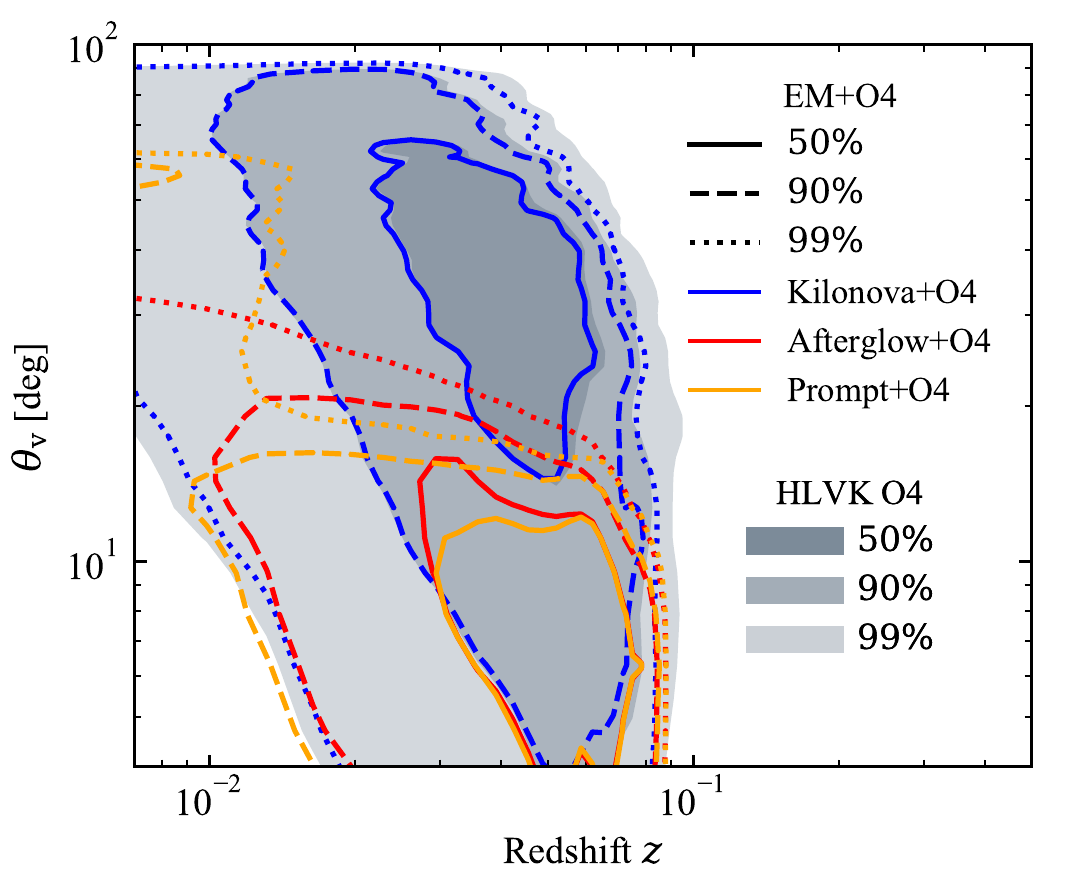}
    \caption{Viewing angle $\theta_\mathrm{v}$ versus redshift $z$ for our NSNS population. The filled grey regions contain $50\%$, $90\%$, $99\%$ of the GW O4-detectable binaries. Solid, dashed and dotted contours contain $50\%$, $90\%$, $99\%$ of the binaries that exceed both the O4 GW SNR$_\mathrm{net}$ limit and any one of the `couterpart search' limits relevant to the particular counterpart considered (blue: KN; red: GRB afterglow; orange: GRB prompt). The corresponding detection rates are reported in Figure \ref{fig:detection}.}
    \label{fig:view_redshift}
\end{figure}

\subsection{Viewing angle versus redshift}
In Figure \ref{fig:view_redshift} we show the distribution of some sub-samples of our population in the viewing angle $\theta_\mathrm{v}$ versus redshift $z$ plane. Grey filled contours refer to HLVK O4-detectable binaries, while empty contours refer to joint GW and EM detectable binaries: in particular the blue, orange and red lines refer to KN+O4, GRB Prompt+O4 and GRB Afterglow+O4 detectable binaries, respectively. The detection rates corresponding to these regions are shown by lines of the same color in Figure \ref{fig:detection} and are reported in Table \ref{tab:det_rates}. The figure clearly shows the weak dependence on redshift for the jet-related emission, whose luminosity is strongly dependent on the viewing angle. Moreover, $90\%$ ($50\%$) of the GRB Prompt+O4 and GRB Afterglow+O4 events have relativistic jets seen under a viewing angle lower than $\sim 15 \, (10) $ degrees. 

\subsection{Detection rate versus detection limit}\label{apx:det_lim}
\begin{figure}
    \centering
    \includegraphics[width=0.9\textwidth]{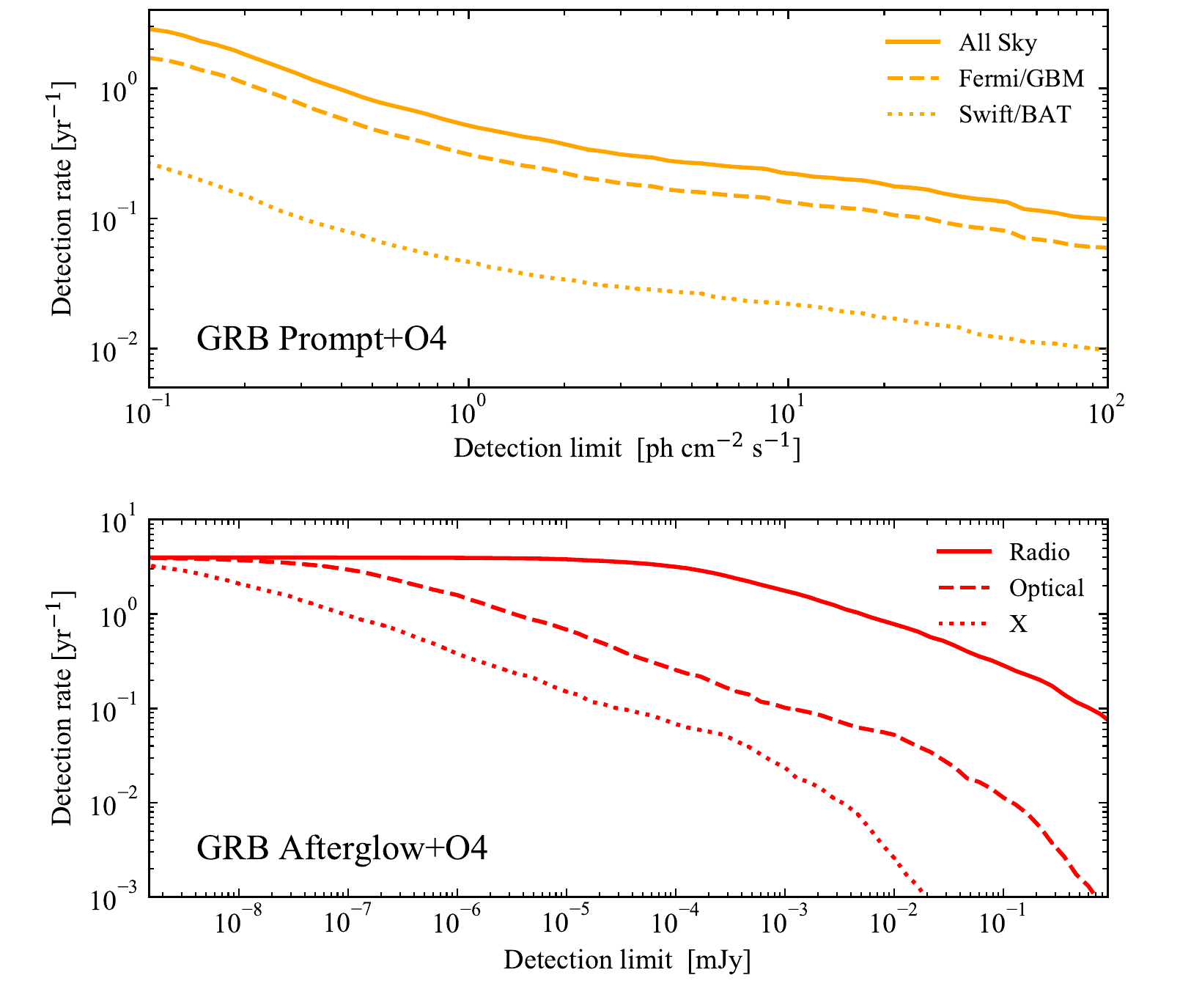}
    \caption{Detection rates as a function of the detection threshold limit for our NSNS population. The upper panel refers to GRB Prompt+O4 detectable binaries. The solid line indicates an all-sky field of view with a $100\%$ duty cycle, the dashed and dotted lines accounts for the \textit{Fermi}/GBM and \textit{Swift}/BAT duty cycle and field of view, respectively. The lower panel refers to GRB Afterglow+O4 detectable binaries. The solid, dashed and dotted lines indicate the radio, optical and X band, respectively.}
    \label{fig:det_lim}
\end{figure}

In section \ref{sec:EMO4} we report the detection rates for joint GW and EM events considering two representative detection limit sets based on the two main scenarios considered in this work. In order to allow the community to explore alternative observing configurations that correspond to different detection limits, we show in Figure \ref{fig:det_lim} the distribution of the detection rates as a function of the detection limit for the GRB Prompt+O4 (upper panel, orange) and GRB Afterglow+O4 (lower panel, red) detectable binaries (for KN\ae, such information is already contained in the right-hand panels of Figures \ref{fig:kn} and \ref{fig:kn_gJ}).  

For the GRB Prompt+O4 detection we show the rates assuming an all-sky field of view and a $100\%$ duty cycle (solid line) and accounting for the duty cycle and field of view of \textit{Fermi}/GBM (dashed line) and \textit{Swift}/BAT (dotted line). The figure shows how the GRB prompt+GW detection rate increases with the prompt emission detector sensitivity: if it were possible to reach photon flux threshold values of $\sim$0.1 ph cm$^{-2}$ s$^{-1}$, the cocoon emission would start to be detected in essentially all jet-launching binaries (this produces the bump in the orange lines at the low-flux-limit end).

For the GRB Afterglow+O4 events, we show individually the rates for the radio (solid), optical (dashed) and X (dotted) bands. The detection limit value at which the curves saturate indicates the sensitivity needed to detect all the GRB Afterglow+O4 events, with a corresponding detection rate of $4.0^{+6.1}_{-3.0}$ $ \mathrm{yr}^{-1}$ (that is the GW O4 detection rate of \searchhlvk\ yr$^{-1}$ times the \fracjet\ fraction of jet-launching system).


\bibliography{references.bib}{}
\bibliographystyle{aasjournal}

\end{document}